\documentclass[12pt]{article}
\usepackage{amsmath,amsthm,amssymb,euscript,epsf,epsfig}
\usepackage{array}
\usepackage{fancybox}
\usepackage[usenames]{color}
\usepackage{amsfonts,bm}
\usepackage{graphicx}
\usepackage{enumerate}

\usepackage[
      colorlinks=true,
      linkcolor=blue,
      urlcolor=blue,
      filecolor=blue,
      citecolor=blue,
      pdfstartview=FitH,
      pdftitle={},
        pdfauthor={},
        pdfsubject={},
        pdfkeywords={},
        pdfpagemode=UseNone,
        bookmarksopen=true
      ]{hyperref}
\usepackage[all]{hypcap}     

\newcommand{\be}{\begin{equation}}
\newcommand{\ee}{\end{equation}}
\newcommand{\bea}{\begin{eqnarray}\displaystyle}
\newcommand{\eea}{\end{eqnarray}}

\newcommand{\nn}{\nonumber}

\makeatletter
\@addtoreset{equation}{section}
\makeatother

\def\one{{\hbox{ 1\kern-.8mm l}}}
\def\zero{{\hbox{ 0\kern-1.5mm 0}}}

\def\cO{ { \cal{O}}}
\def\g{ \gamma} 
\def\b{ \beta } 
\def\a{ \alpha } 
\def\d{ \partial } 
\def\zb{{\ov z}}

\def\mC{ \mathbb{C} } 
  
\def\mR{ \mathbb{R}}

\def\s{ \sigma} 
\def\th{\theta}

 \def\cB{{\cal B}} \def\cC{{\cal C}}
\def\cD{{\cal D}} \def\cE{{\cal E}} 
 \def\cH{{\cal H}} 
  \def\cL{{\cal L}}
\def\cM{{\cal M}} \def\cN{{\cal N}} \def\cO{{\cal O}}
\def\cP{{\cal P}}  
\def\cS{{\cal S}}  
  
 \def\cZ{{\cal Z}}

\definecolor{orange}{rgb}{1,0.5,0}

\def\mC{ \mathbb{C} }

\def\mR{ \mathbb{R}} 

\def\vmn{ V^{\otimes m } \otimes \bar V^{\otimes n } }
\newcommand{\bra}[1]{{\langle {#1} |\,}}
\newcommand{\ket}[1]{{\,| {#1} \rangle}}
\newcommand{\braket}[2]{\ensuremath{\langle #1 | #2 \rangle}}
\newcommand{\Braket}[2]{\ensuremath{\langle\, #1 \,|\, #2 \,\rangle}}

\newcommand{\sq}{\square}

\newcommand{\floor}[1]{\left\lfloor #1 \right\rfloor}
\newcommand{\ceil}[1]{\left\lceil #1 \right\rceil}

\newcommand{\Lb}{\ensuremath{\bar{L}}}

\newcommand{\dbyd}[1]{\ensuremath{ \frac{\d}{\d {#1}}}}

\newcommand{\Zd}{\ensuremath{ Z^{\dagger}}}

\newcommand{\Ad}{\ensuremath{ A^{\dagger}}}
\newcommand{\Bd}{\ensuremath{ B^{\dagger}}}
\newcommand{\Ud}{\ensuremath{ U^{\dagger}}}
\newcommand{\Td}{\ensuremath{ T^{\dagger}}}

\newcommand{\T}[3]{\ensuremath{ #1{}^{#2}_{\phantom{#2} \! #3}}}		
\newcommand{\de}[2]{\ensuremath{ \T{\delta}{#1}{#2}}}

\def\vac{|0\rangle}
\newcommand{\tr}{\operatorname{tr}}

\def\ha{\frac{1}{2}}

\def\a{\alpha}
\def\b{\beta}

\def\g{\gamma}

\def\f{\phi}
\def\F{\Phi}

\def\o{\omega}

\def\s{\sigma}

\def\l{\lambda}
\def\L{\Lambda}
\def\p{\psi}

\def\d{\partial}

\def\dag{\dagger}

\def\th{\theta}

\def\zb{{\ov z}}

\def\ov{\overline}

\def\R{\mathbb{R}}

\def\half{\frac{1}{2}}
\def\li{\lambda_i}

\def\eq#1{(\ref{#1})}

\begin{document}

{}~
{}~
\hbox{QMUL-PH-09-20}
\break

\vskip 1cm

\centerline{
{\LARGE \bf  Free particles from Brauer algebras}} 
\vskip .4cm
\centerline{
{\LARGE \bf  in complex matrix models }}
                                                       
\vskip 1.2cm

\centerline{ {\large \bf Yusuke Kimura}\footnote{y.kimura@qmul.ac.uk},~
  {\large \bf Sanjaye Ramgoolam}\footnote{s.ramgoolam@qmul.ac.uk}    
{ \bf and} {  {\large \bf David Turton}\footnote{d.j.turton@qmul.ac.uk}  } }  
\vskip 1.2cm
\begin{center}
{\large Queen Mary University of London\\
Centre for Research in String Theory \\ 
Department of Physics\\
Mile End Road\\
London E1 4NS UK\\
}
\end{center}

\vspace*{5.0ex}

\centerline{\bf Abstract} \bigskip

The gauge invariant degrees of freedom 
of matrix models  based on an $ N \times N $
complex matrix, with $U(N)$ gauge symmetry, 
contain hidden free particle structures. 
These are exhibited using 
triangular matrix variables via the Schur decomposition. 
The Brauer algebra basis for complex matrix models developed 
earlier is useful in projecting to  a sector which matches
 the state counting of 
 $N$ free fermions on 
a circle. The Brauer algebra projection  is characterized 
by  the vanishing of a scale invariant laplacian constructed from 
the complex matrix. The special case 
of $N=2$ is studied in detail:  the ring of gauge invariant 
functions as well as a ring  of scale and 
gauge invariant differential operators are 
characterized completely.  The orthonormal basis of wavefunctions
in this special case is completely characterized by 
a set of five commuting Hamiltonians, which display free particle structures.
Applications to the reduced matrix quantum mechanics coming from 
radial quantization in $\cN=4$ SYM are described. 
We propose  that the string dual of the  complex matrix harmonic oscillator 
quantum mechanics  has an interpretation in terms of strings and branes 
in  $2+1 $ dimensions.

\thispagestyle{empty}
\vfill

\eject

\tableofcontents

\section{Introduction} 
\setcounter{footnote}{0}

\subsection{Background and motivations } 

There is a class of Gaussian matrix models in $D$ spacetime dimensions\footnote{By `matrix models in $D$ spacetime dimensions' we include random matrix models $(D=0)$, matrix quantum mechanics $(D=1)$ or field theories. By `Gaussian' we mean a quadratic Lagrangian.} $x^{\mu}$ 
where the two-point function of the matrix $Z( x^{\mu})$, up to a trivial spacetime 
dependence, is 
\bea\label{basic2pt}  
 \langle \T{Z}{i}{j} \, \T{\Zd}{k}{l} \rangle 
 =  \T{\delta}{i}{l} \T{\delta}{k}{j}
\eea 
and where there is an adjoint $U(N)$ action
\bea\label{Uact} 
Z \rightarrow g Z g^{\dagger},  \qquad\qquad g \in U(N).
\eea 
Our main interest is in $D=4$, $\cN=4$ superconformal Yang-Mills (SYM) with gauge group $U(N)$ at zero coupling and its dimensional reduction on $S^3$ to a $D=1$ matrix harmonic oscillator quantum mechanics.  Some of our results apply more generally and in particular we make connections to the $D=0$ Gaussian complex matrix model considered by Ginibre \cite{Ginibre}.

$\cN=4$ super Yang-Mills contains three complex scalar fields in the adjoint of the gauge group. The states built from holomorphic functions in one of these fields, say $Z$, comprise the half-BPS sector in which the two-point function is diagonalised by the Schur polynomials $\chi_R(Z)$ \cite{Corley:2001zk}:
\bea 
\langle \chi_S ( Z^{\dagger } ) \chi_R ( Z ) \rangle \propto \delta_{RS} 
\eea  
The Schur polynomials were identified as gauge theory duals of giant 
gravitons \cite{McGreevy:2000cw}, generalizing the proposal of \cite{Balasubramanian:2001nh} 
which associated determinants and sub-determinants to giant gravitons expanding in
the $S^5$ of the $AdS_5 \times S^5$ dual of $\cN=4$ SYM. This is manifestation of the
`stringy exclusion principle' \cite{Maldacena:1998bw,Jevicki:1998rr}.

The Schur polynomial basis allowed the identification 
of gauge theory duals for multiple giants as well as 
giants expanding in the $ AdS_5$ directions. 
Extensive evidence for the proposed map has been found, see for 
example the review \cite{Koch:2009gq}. 
It was also observed that Schur Polynomials were related to wavefunctions of free fermions
which are the eigenvalues of the matrix $Z$ \cite{Corley:2001zk,Berenstein:2004kk}.
The free fermions were subsequently identified 
as arising from supergravity solutions \cite{Lin:2004nb}. 
Non-renormalization theorems \cite{Eden:2000qp,Eden:2000gg} allow a comparison of 
zero coupling to strong coupling; the diagonalisation of the two-point function 
was a crucial step enabling this comparison. \newline

Motivated by the goal of understanding non-supersymmetric 
sectors of the $AdS_5 / CFT_4$ gauge-string duality, \cite{Kimura:2007wy} 
undertook the study of gauge invariant non-holomorphic functions of $Z$.
Two-point functions of operators which are polynomials 
of degree $m$ in $Z$ and degree $n$ in $\Zd$ in the $U(N)$ theory 
were diagonalised
 in terms of a {\it Brauer basis} which was constructed systematically using 
 the representation theory of the Brauer algebra $B_N(m,n)$, previously studied in
\cite{Stembridge:1987,Koike:1989,Benkart:1992,Halverson:1996}.
This non-holomorphic sector is arguably the simplest non-supersymmetric sector, 
yet contains many of the subtleties of 
multi-matrix combinatorics, since in general $Z$ does not commute
with its conjugate transpose $\Zd$. \newline

The interest in a detailed study at zero Yang-Mills coupling is two-fold.
Firstly, the strong form of the Maldacena conjecture \cite{Maldacena:1997re} 
implies that there is a string theory dual to zero coupling Yang-Mills theory.
At zero coupling, the sector containing only $Z, \Zd$ forms a consistent truncation and
we may ask whether there is a string dual of the quantum mechanics of the resulting complex matrix quantum mechanics.

Secondly, while there is undoubtedly less control in comparing zero coupling 
with strong coupling and thus to semiclassical brane solutions and supergravity 
solutions it is possible that some qualitative features uncovered might, in appropriate 
large charge regimes, survive in the strong coupling limit. This line of reasoning is 
used, for example, in \cite{Balasubramanian:2007bs} where black hole 
entropy is investigated from the counting of gauge invariant operators at zero coupling. 
For some earlier works on complex matrix models, see for example
\cite{Kostov:1997bn,Kristjansen:2002bb,Akemann:2002vy,Alexandrov:2009gn}.

In this paper we always work at finite $N$. There is an important distinction between
\be
N \ge m + n \qquad \mathrm{and} \qquad  N< m + n ~.
\ee
The condition $N \ge m + n $ may be read as a condition that $N$ be larger than the lengths of operators one wishes to discuss; for example taking the planar limit $N \to \infty$ achieves this trivially. For fixed finite $N$, this is the regime in which lengths of operators are less than or equal to $N$.
The opposite regime $N< m+n$ is relevant to studies of heavy operators in $\cN=4$ super Yang-Mills such as conjectured duals of multi-branes and black holes in $AdS_5 \times S^5$.

The representation theory of Brauer algebras, and thus the 
construction of the Brauer basis, is well understood for $N \ge m + n$
however there are interesting subtleties for $N < m + n $ (see for example \cite{Cox:2007}). \newline

In single hermitian or unitary (or more generally, normal) matrix models, 
the unitary group action \eq{Uact} is sufficient to diagonalise the matrix,
leading to a free particle description in terms of the $N$ eigenvalues as we shall review.
In an unrestricted complex matrix model, this is not sufficient to 
diagonalise $Z$; a generic complex matrix may at best be put into triangular form.

More precisely, using the Schur decomposition $ Z =  U T U^{\dagger}$ where $T$ is 
upper triangular, the space $gl(N;\mathbb{C})$ may be decomposed into a parameter space of 
inequivalent orbits $\cM_{ N }$ and the orbits of the 
 $U(N)$ action. $\cM_{ N }$ has real dimension $N^2+1$ and is a fibration over 
the symmetric product $Sym^N ( \mC )$:
\bea 
 \hskip1in   && ~~ \cM_N \cr 
  \hskip1in  && ~~~   \downarrow \cr 
 \hskip1in   && Sym^N ( \mC ) = { \mC^N / S_N } 
\eea 
The eigenvalues are however coupled to the off-diagonal triangular entries and so cannot 
represent positions of free particles.

We shall show that free particles arise in a non-trivial way by exploiting a map
identified in \cite{Kimura:2007wy} between the $k=0$ sector (to be defined later)
of the Brauer basis and a unitary matrix model, providing in turn a map to $N$ free fermions on a circle. 
In this paper we will give evidence for the following conjecture: that these $N$ 
free fermions of the $k=0$ sector can be constructed 
from degrees of freedom which are composed of eigenvalues
as well as off-diagonal elements of the matrix $Z$. We also observe a different emergence of free 
particles in the $m=n=k$ sector\footnote{As this paper was being written up,
 we became aware of \cite{rodrigues} which studies this sector
and the associated free fermions using a matrix polar decomposition.}.
While we start with the gauge invariant sector of a Gaussian complex matrix 
model, which is a system of   $N^2$ particles 
{\it constrained } by the gauge invariance condition, 
the emergent particles are $N$ free  fermions without constraints.

Much of our work has been carried out $N=2$ as this allows 
explicit calculations. Many of our $N=2$ results extend to general $N$ as 
discussed in Section \ref{genNFacts}; in particular the key point of 
free particles emerging from a Matrix model 
from degrees of freedom \textit{beyond eigenvalues} is valid for any $N$.   \newline

\subsection{Outline of paper}

The structure of the paper is as follows. Section \ref{sec:review} reviews
the emergence of free particles from the eigenvalues 
of hermitian and unitary matrix models and explains the 
new features arising for a complex matrix model. 
 Section \ref{sec:orbitsparams} reviews the Schur decomposition,  
gauged matrix quantum mechanics and describes the orbits over $\cM_N$ in general 
and over $\cM_2$ in detail.
In Section \ref{truncN=2} we describe the ring of  
functions on $\cM_2$, which come from gauge 
invariant polynomials on $gl(2, \mC )$.  

In Section \ref{casfuns} we describe the ring of Casimir 
operators studied in \cite{Kimura:2008ac},
we present computational results on the Brauer basis counting at $N=2$
and state a conjecture for the complete solution.
The conjectured counting can be elegantly described in terms of five 
integer labels and is the first main result of this paper.

We derive explicit expressions for 
the Casimir operators as differential operators on $\cM_{2}$ and 
express the integer labels as functions of the Casimirs. 
We define free particle momentum operators and 
express these operators as functions of differential 
operators on $\cM_{2}$. Amongst these operators are 
the conjectured $k=0$ sector free fermion momenta on a circle.
This is the second main result of this paper. 

Section \ref{genNFacts} presents a conjecture that 
the $k=0$ sector is the kernel of a scale-invariant laplacian
on $\cM_{N}$. We give expressions for a class of 
three-point functions of operators in the $k=0$ sector 
in terms of unitary matrix integrals, which provides 
further evidence for the conjectured equivalence to 
$N$ free fermions on a circle. We extend some remarks 
on the counting of states at $N=2$ to higher $N$. 

We also find free particle structures in the $m=n=k$ sector, 
by observing that this sector consists of  multi-traces of the combination
$ Z^{\dagger}Z$. We show that this sector may be identified
with the kernel of a differential operator on $ \cM_{N}$.

Most of the material in Sections \ref{sec:orbitsparams}-\ref{genNFacts}
refers to the properties related to invariants under the $U(N)$ action \eq{Uact} and as such 
is relevant to general complex matrix models with $U(N)$ symmetry.
Section \ref{sec:SHO} deals specifically with the matrix quantum 
mechanics obtained by dimensional reduction of SYM and draws a connection 
with Ginibre's $D=0$ matrix model.
Using the higher conserved charges to define 
new Hamiltonians, and the expressions from Section \ref{casfuns}, 
we find non-holomorphic generalizations of the Calogero-Sutherland models 
at special couplings.  Technical details are presented in the Appendices.

\vspace{0.5cm}

\section{Review of free particles in Matrix models } \label{sec:review}

We briefly review examples of hermitian and unitary matrix models which arise in the context of string theory, in particular string theory in two dimensions. This review is not intended to be complete in any sense but rather to provide the reader with context for the current work.

\subsection{Hermitian matrix quantum mechanics} \label{sec:HMM}

Let us consider the Gaussian hermitian matrix model defined by the Lagrangian
\be
\cL =  \tr \left( \frac{1}{2}\dot{\F}^2 - \frac{1}{2}\F^2 \right)
\ee
which is invariant under the $U(N)$ action
\bea\label{UactH} 
\F \rightarrow g \F g^{\dagger}~,  \qquad\qquad g \in U(N)~.
\eea   
We follow the treatment in \cite{Klebanov:1991qa,Ginsparg:1993is,Alexandrov:2003ut} restricting attention to the theory with quadratic potential.  
The Hamiltonian of this model is 
\begin{eqnarray}
H &=&  \tr \left(
  -  \frac{1}{2}\frac{\partial^{2}
}{\partial \F \partial \F }
+\frac{1}{2} \F^2  \right) ~.
\end{eqnarray}
Introducing the annihilation and creation operators
\begin{eqnarray}
A=\frac{1}{\sqrt{2}}\left(\F
+\frac{\partial}{\partial \F}\right) 
\quad
A^{\dagger}=
\frac{1}{\sqrt{2}}\left(\F
-\frac{\partial}{\partial \F}\right)
\end{eqnarray}
and using the usual convention for matrix indices
\be \label{eq:matrixderiv}
\left( \dbyd{\F} \right)^{\!\!i}_{j} = \dbyd{\T{\F}{j}{i}}
\ee
we have $[\T{A}{i}{j},\T{A^{\dagger}}{k}{l}]=\T{\delta}{k}{j} \T{\delta}{i}{l}$
and the Hamiltonian can be rewritten as 
\begin{eqnarray}
H=\tr (A^{\dagger}A)+\frac{N^2}{2} ~.
\end{eqnarray}
The ground state has energy  $\frac{N^2}{2}$ and its wavefunction is
\begin{eqnarray}
\Phi_{0}=\langle \F| 0\rangle =e^{-\frac{1}{2}\tr \F^{2}}~.
\end{eqnarray}
$U(N)$ singlet excited states are obtained by acting on $\Phi_{0}$ 
with $U(N)$ invariant functions of $\Ad$, 
or by absorbing factors of $\sqrt{2}$, multiplying by $U(N)$ invariant functions of $\F$. 
A basis for such functions is given by the Schur polynomials, which are polynomials of degree $n$ labelled by a representation $R$ of $S_n$,
\be
\chi_R(\Phi) = \sum_{\s \in S_n} \chi_R(\s) \T{\Phi}{i_1}{i_{\s_1}} \cdots \T{\Phi}{i_n}{i_{\s_n}}~,
\ee
where $\chi_R(\s)$ is the character of $\s$ in the representation $R$. 
The associated wavefunction
\be
\Psi_R = \chi_R(\Phi) e^{-\frac{1}{2}\tr \F^{2}}
\ee
has energy $\frac{N^2}{2}+n$.

A hermitian matrix $\F$ may be decomposed as 
\be \label{eq:Hdecom}
\F = U \L U^{\dag}, \qquad \L = diag(\l_1,\ldots,\l_N)~, \qquad U \in U(N)
\ee
under which we get
\begin{eqnarray}
\tr(\dot{\Phi}^{2})=\tr(\dot{\Lambda}^{2})+\tr[\Lambda,U^{\dagger}\dot{U}]^{2}~.
\end{eqnarray}
The anti-hermitian matrix $U^{\dagger}\dot{U}$ may be expanded in generators of $U(N)$ as 
\begin{eqnarray*}
U^{\dagger}\dot{U}=\sum_{i}\alpha_{i}H_{i}
+\frac{i}{\sqrt{2}}\sum_{j<k}(\dot{\alpha}_{jk}T_{jk}+\dot{\beta}_{jk}\widetilde{T}_{jk})
\end{eqnarray*}
where $H_i$ are the diagonal generators of the Cartan subalgebra,
$T_{jk}$ is the matrix $M$ such that $M_{jk}=M_{kj}=1$ and
all other entries are 0, and $\widetilde T_{ij}$ is the matrix $M$
such that $M_{ij}=-M_{ji}=-i$ and all other entries are 0. This gives
\begin{eqnarray*}
\tr\,[\Lambda,U^{\dagger}\dot{U}]^{2}=
\sum_{i<j}(\lambda_{i}-\lambda_{j})^{2}
(\dot{\alpha}_{ij}^{2}+\dot{\beta}_{ij}^{2})
\end{eqnarray*}
and so the Lagrangian becomes
\be
L=\sum_i\left(\ha \dot\li^2+\ha \li^2 \right)
+\half\sum_{i<j} (\li-\l_j)^2(
\dot\a_{ij}^2 +\dot\b_{ij}^2)~.
\ee
Under the transformation \eq{UactH} the measure becomes
\be
{\cal D} \Phi ={\cal D} \Omega\prod_i d\l_i \Delta^2(\l)
\ee
where $\Delta(\l)$~is the Vandermonde determinant
$\prod_{i<j}(\l_i-\l_j)$, the kinetic term for the eigenvalues becomes
\bea
 -{1 \over 2} \sum_i
{1\over\Delta^2(\l)}{d\over d \lambda_i}
\Delta^2(\l){d\over d \lambda_i} &=& -{1 \over 2 \Delta(\lambda )} \sum_i{d^2\over {d \lambda_i}^2}
\Delta(\l)\
\eea
and so the Hamiltonian is 
\begin{eqnarray}
H 
&=& \frac{1}{2}\sum_{i}\left(
-\frac{1}{\Delta(\lambda)}
\frac{\partial^{2}}{\partial \lambda_{i}^{2}}
\Delta(\lambda)
+\lambda_{i}^2 \right)
-\frac{1}{2}\sum_{i<j}
\frac{1}{
(\lambda_{i}-\lambda_{j})^{2}}
\left(
\frac{\partial^{2}}{\partial \alpha_{ij}^{2}}
+
\frac{\partial^{2}}{\partial \beta_{ij}^{2}}
\right)~.
\end{eqnarray}
Wavefunctions which are singlet under \eq{UactH} are symmetric functions of the eigenvalues, $\chi_{sym}(\l)$. On these wavefunctions the Hamiltonian simplifies to 
\bea
H&=& \frac{1}{2}\sum_{i}\left(
-\frac{1}{\Delta(\lambda)}
\frac{\partial^{2}}{\partial \lambda_{i}^{2}}
\Delta(\lambda)
+\lambda_{i}^2 \right)~.
\eea
One may simplify further the analysis by defining the antisymmetric wavefunction
\be
\Psi^f(\l) =\Delta(\l) \chi_{sym}(\l)
\ee
and the modified Hamiltonian
\be
H^f ~=~  \Delta(\l) H \frac{1}{\Delta(\l)} ~=~ {1 \over 2 } \sum_{i} \left( - {d^2\over {d \lambda_i}^2} + \lambda_i^2 \right)
\ee
which is a sum of one particle harmonic oscillator Hamiltonians. Then $H^f$ has eigenstates $\Psi^f(\l)$ with the same eigenvalues as $H$:
\bea
H \Psi(\l)  &=& E \,\Psi(\l) \\
\Rightarrow H^f \Psi^f(\l)  &=& E \,  \Psi^f(\l)~.
\eea
The ground state wavefunction of $H^f$ is 
\be
\Psi^f_0 = \Delta e^{- \ha \tr \Phi^2 }~,
\ee
excited states are given by Slater determinants
\be
\Psi^f_{\vec \cE} ~=~ \det_{i,j} \l_i^{\cE_j} e^{- \ha \tr \Phi^2 } ~=~ \Delta(\l) \Psi_R(U)
\ee
and so the $U(N)$ singlet sector is equivalent to $N$ non-interacting fermions in a harmonic oscillator potential,
where the fermion energies $\cE_i$ are related to the integer row lengths $r_i$ of $R$ by
\be \label{eq:energies-SHO}
\cE_i  =  r_i  +  ( N - i ) ~.
\ee

\subsection{Unitary matrix quantum mechanics} \label{sec:UMM}

We next review the unitary matrix quantum mechanics which arises in the study of two-dimensional Yang-Mills, which is given by the Hamiltonian \cite{Douglas:1993xv,Douglas:1993wy}: 
\be 
H ~=~ \tr \left ( U \dbyd{U} \right)^2 ~=~ \sum_{a} E^a E^a
\ee
where $E^a$ generates left rotations of $U$:
\be
E^a = \tr t^a  U \dbyd{U}
\ee
The form of $H$ means that acting on a wavefunction which is a matrix element of an irreducible representation $R$,
\be
(\psi_R)_{ij}(U) = D^R_{ij}(U)
\ee
it measures the quadratic Casimir of the representation $R$, 
\be
H\psi_R(U) = C_2(R) D^R_{ij}(U)~.
\ee
Representations are classified by their characters, the Schur polynomials
\be
\chi_R(U) = \tr D^R(U)
\ee
which form an orthonormal basis for wavefunctions invariant under the $U(N)$ action
\be \label{eq:adjactionU}
U \to g U g^{\dag}, \qquad g \in U(N)~.
\ee
This may be used to express any unitary matrix $U$ as
\be \label{eq:adjactionU2}
U = g D g^{\dag}, \qquad D = diag(e^{i\th_1},\ldots,e^{i\th_N}), \qquad g \in U(N)~.
\ee
On functions invariant under (\ref{eq:adjactionU}), performing the change of variables (\ref{eq:adjactionU2}) the Hamiltonian becomes \cite{Douglas:1993xv}:
\be
H ~=~ -\sum_i \left[ {1\over\tilde\Delta}{d^2\over
d\theta_i^2}
\tilde\Delta \right]  - \frac{1}{12} N (N^2-1)
\ee
where denoting the eigenvalues by $u_i = e^{i\th_i}$,
\be
\tilde\Delta ~=~ \prod_{i<j}\sin{\theta_i-\theta_j\over 2} 
~=~ \frac{\Delta(u)}{\prod_i u_i^{\frac{N-1}{2}}} ~=~ \frac{\Delta(u)}{(\det U)^{\frac{N-1}{2}}}
\ee
and where
\be
\Delta(u)=\prod_{i<j}(u_i-u_j)~.
\ee 
Absorbing $\tilde\Delta$ into the wavefunctions and the Hamiltonian, 
\be \label{eq:H_f-UMM}
\psi_f = \tilde\Delta\psi ~, \qquad H_f ~=~ \tilde\Delta H \frac{1}{\tilde\Delta} ~=~ \sum_i \dbyd{\th_i^2} - \frac{1}{12} N (N^2-1)
\ee
the wavefunctions become antisymmetric under exchange of any pair $\th_i \leftrightarrow \th_j$.
The one-particle wavefunctions with quantized momentum $p$ are $\psi_p = e^{ip \th}$ and the Slater determinants
\be
\psi_{\vec p} ~=~ \det_{i,j} u_i^{p_j}
\ee
are eigenfunctions of $H_f$ with energy $E = \sum_i p_i^2 - N(N^2-1)/12$, so the sector of this theory invariant under (\ref{eq:adjactionU}) is equivalent to a theory of $N$ free fermions on a circle. The ground state has fermion with momenta distributed symmetrically about $n=0$, and
energy zero, so the Fermi energy is $n_F=\frac{N-1}{2}$ and there are Fermi surfaces at $\pm n_F$. The Slater determinants are related to the Schur polynomials via
\be
\psi_{\vec p} ~=~ \Delta(u) \chi_R(U)
\ee
where the momenta $p_i$ are related to the integer row lengths $r_i$ of $R$ by
\be \label{eq:energies-circle}
p_i  =  r_i  +  ( n_F + 1 - i ) ~.
\ee

\subsection{Complex matrix models} \label{sec:CMM}

Previous studies of complex matrix models have centred on models in which there is enough symmetry to diagonalise the matrix. This can be achieved by studying a normal matrix ($[Z, \Zd] = 0$) with $U(N)$ symmetry (see e.g. \cite{Chau:1991gj,Chau:1997pr})
\be \label{eq:adjaction}
Z \to g Z g^{\dag}~, \qquad g \in U(N)
\ee
or by studying an unrestricted complex $Z$ with $U(N) \times U(N)$ symmetry (see e.g.~\cite{Klebanov:2003wg})
\be
Z \to g Z h^{\dag}~, \qquad g, h \in U(N)~.
\ee
In this paper, motivated by gauge-gravity duality we study an unrestricted complex matrix $Z$ with a single $U(N)$ symmetry (\ref{eq:adjaction}). This requires us to go beyond an eigenvlue description and take into account off-diagonal degrees of freedom. \newline

Due to the off-diagonal degrees of freedom we do not expect a 
straightforward transformation to a description in terms of free
 particles for complex matrix models with unitary symmetry. 
Nevertheless, our investigations of the $k=0$ sector of the Brauer 
basis indicate that the free particles on a circle
of the Unitary matrix model
can be constructed from the degrees of freedom of 
the complex matrix model. These free particles on a circle 
are emergent degrees of freedom arising from 
eigenalues and off-diagonal elements constrained 
by equations which define the $k=0$ sector.

In passing we note that the Gaussian complex matrix model with $U(N)$ symmetry (\ref{eq:adjaction}) may be written as a two-Hermitian matrix model \cite{Itzykson:1979fi} using  
\be
X = \frac{1}{2} \left( Z + \bar{Z} \right) , \qquad  Y = - \frac{i}{2} \left( Z - \bar{Z} \right) 
\ee
where $\bar{Z}$ denotes complex conjugate of $Z$.
Studies of the same model in terms of two hermitian matrices 
are done in \cite{Rodrigues:2005ec,Rodrigues:2008uh}.

\section{Orbits and parameter spaces}\label{sec:orbitsparams} 

The relation between  $ gl( N ,  \mC )  $, the space of 
complex matrices $Z $ and the space $ \cM_N$, of 
orbits under the adjoint action (\ref{Uact}),  
is given by the Schur decomposition.

\subsection{Orbits and the structure of $\cM_N$} \label{sec:schur}

Schur's decomposition (see e.g.~\cite{Meyer:2000Schur})
allows one to write any complex matrix $Z$ as 
\be\label{ZUT} 
Z = U T \Ud
\ee
where $U \in U(N)$ and $T$ is upper triangular.
It has been used previously in the context of the complex matrix model 
in \cite{Mehta,Takayama:2005yq}.  The eigenvalues $z_i$ of $Z$ become the diagonal 
entries (and hence the eigenvalues) of $T$. 
There are also off-diagonal elements $t_{ij}$ for 
$ i < j $. The equation  (\ref{ZUT}) can be viewed as describing a
map from the pair $( U, T)$ to complex matrices. 
The map is onto, but not one-to-one. Pairs $( U, T ) $ and 
$(  e^{ i \theta } U , T ) $  describe the same $Z$. 
There is a $U(1)^{N}$ action 
\bea
U &\to U' =& UH  , \qquad H  = diag(e^{i\theta_1}, \ldots,e^{i\theta_N})  \nn \\
T &\to T' =& H^{\dag} T H  \label{eq:u1N}
\eea
which leaves $Z$ unchanged. The diagonal $ e^{i \theta} $ acts trivially 
on $ T$ but the $U(1)^{N-1}$ part defined by $ \sum \theta_j =0 $ 
mixes non-trivially with the angles in $T$. 

We can  parameterize 
the coset $ U(N)/U(1)^N $ using the variable $ L$
and decomposing  $ U = L H $ (as for example in \cite{Castellani:1999fz}) 
leading to 
\be \label{ZLT}
Z  ~~ = ~~  L ( H  T H^{\dagger} ) L^{\dagger}  ~~ = ~~  L  \tilde T L^{\dagger} 
\ee  
where $ \tilde T \equiv  H  T H^{\dagger}$. 
It is also convenient to use the $U(1)^{N-1}$ part of (\ref{eq:u1N}) to set the $N-1$ 
entries on the superdiagonal of $T$ (namely $t_{j,j+1}$) to be real, and to use $(U,T)$. 

There is also the freedom, for fixed $Z$, to rearrange the eigenvalues in any order on the diagonal of $T$ by altering $U$. This freedom exists because there is a Schur decomposition for each possible ordering of eigenvalues on the diagonal of $T$. Given 
\be
Z  ~~ = ~~  U_1  T_1  U_1^{\dagger}  ~~ = ~~  U_2  T_2  U_2^{\dagger}
\ee 
where $T_1$ and $T_2$ have different orderings of diagonal entries, we have
\be \label{T2T1}
T_2 ~~ = ~~ \left( U_2^{\dag} U_1 \right)  T_1  \left( U_2^{\dag} U_1 \right)^{\dag}  ~~ = ~~  U_{12}  T_1  U_{12}^{\dagger}
\ee
where  $U_{12} \equiv  U_2^{\dag} U_1$.

We have thus derived the construction mentioned in the introduction of $\cM_{ N }$ as a fibration over 
the symmetric product $Sym^N ( \mC )$:
\bea 
 \hskip1in  && ~~ \cM_N \cr 
  \hskip1in && ~~~   \downarrow \cr 
 \hskip1in      && Sym^N ( \mC ) = { \mC^N / S_N } ~.
\eea 
The set of eigenvalues $ z_1 , z_2 , \ldots , z_N $ of $Z$ 
modulo permutations in $S_N$ forms the space $Sym^N ( \mC )$.
Local coordinates on the fibre of $ \cM_N$ 
over $Sym^N ( \mC )$ are obtained from the  upper triangular elements 
$t_{ij}$, with $i < j $,  appearing in $T$. \newline

Functions of degree $n$ on $\mR^N/S_N$ and  natural 
inner products on the space of functions, which are expressible in terms of 
integrals, are organised by the 
symmetric group $S_n$.  Since $n$ can be arbitrarily large, 
we may say that $S_{\infty}$, defined as an inductive limit 
from finite symmetric groups (see e.g.~\cite{Olshanski:2003}), is the symmetry organising 
the space of functions on $\mR^N/S_N$.   In the case 
of  $\cM_N$ there is an infinite-dimensional underlying 
Brauer algebra constructed as a limit of finite algebras
$ B_N( m ,n ) $. \newline

The space of $N \times N$ complex matrices $gl( N , \mC )$ consists of 
orbits generated by the $U(N)$ action $ Z \rightarrow U Z U^{\dagger} $. 
Due to the trivial $U(1)$ action the real dimension of the parameter space of orbits $\cM_{N}$ is 
$N^2+1 = 2N^2 - ( N^2 -1 )$.

This suggests that the  number of generators of 
ring of functions on $ \cM_N$ should be $ N^2 +1$.
This works in a straightforward way at $N=2$, but 
in a nontrivial way at $ N=3$. We will come back to 
this in Section \ref{genNFacts}. 

Local coordinates on $\cM_{N}$ are given by
$z_i$ and variables $t_{ij}$. At generic $z_i , t_{ij} $ the orbits are 
topologically $U(N)/U(1) =  SU(N) / Z_N$. 
At $t_{ij} = 0 $, the parameter space $\cM_{N}$ becomes $ Sym^N ( \mC ) $. 
The orbit is then  generically $ SU(N)/ U(1)^{N-1}$.  
Note that, when $U(N)$ acts on its  Lie algebra, the adjoint 
orbits are always K\"{a}hler (and hence even dimensional) \cite{Kirillov}. This is 
no longer the case for orbits in the complexified Lie algebra
$gl(N, \mC )$.

\subsection{Differential Gauss's law} \label{sec:GaugedMM}

Dimensional reduction of $ \cN=4$ SYM onto $\mR_t \times S^3$ yields a $U(N)$ gauged matrix quantum mechanics 
involving a complex matrix $Z(t)$ in the adjoint coupled to a gauge field $A_0(t)$. The action takes the form
\be
\cS = \int{ dt \, \tr \bigg(  D_0 Z (D_0 Z)^{\dag} - Z \Zd  \bigg) }
\ee
where $D_0 Z = \d_0 Z + i [A_0 , Z]$. 

Using the above change of variables (\ref{ZUT}), (\ref{ZLT}) we may derive an expression for the 
$1$-form on $ gl(N; \mC ) $ 
\bea \label{dZLtT}  
dZ = U ( d T + [ \omega , T ] ) U^{\dagger} 
=  L ( d\tilde T + [ V   , \tilde T  ] )L^{\dagger} 
\eea 
where $ \omega = U^{\dagger} dU $ and  $ V  = L^{\dagger} dL $. 
This allows us to write the line element $ \tr ( dZ dZ^{\dagger} ) $ 
in terms of the structure constants of the Lie algebra, 
with a choice of decomposition into coset and sub-algebra. 

As an aside, it is interesting to note that, as a consequence of 
(\ref{dZLtT}) we can  write a quantum mechanics theory with 
$ U(N)$ global symmetry 
\bea 
\cS = \int dt \,  \tr  \Big( \d_0 Z \, \d_0 Z^{\dagger} \Big)
  = \int dt \,  \tr   \left( \d_0 \tilde T + [ V   , \tilde T  ] \right) \left( \d_0 \tilde T^{\dagger} + [ V , \tilde T^{\dagger} ] \right) 
\eea  
as a quantum mechanics with 
gauged $U(1)^N$ symmetry and charged matter fields $ \tilde T$
where the one form $V$ on the coset  couples as a gauge field. 
The gauge symmetry is $ \tilde T \rightarrow h \tilde T h^{-1} $ 
and $ V (y) \rightarrow h V h^{-1} + h \d_0 h^{-1}$, under which 
$( \d_0 \tilde T + [ V   , \tilde T  ] )$ transforms covariantly and the 
action is invariant. 

We next review remarks contained in \cite{Kimura:2008ac} and introduce notation we shall use later. A convenient gauge fixing choice is to set $A_0 = 0$. The equation of motion for $A_0$ must still be imposed, leading to Gauss's Law:
\be \label{eq:gauss1}
\Zd \dot Z + Z \dot{Z}^{\dag} - \dot Z \Zd - \dot{Z}^{\dag} Z = 0 ~.
\ee
Upon canonical quantization this leads to the differential form of Gauss's Law, 
which can be written as 
\be\label{Gauss} 
G = G_1 + G_2 + G_3 + G_4 = 0
\ee
where $G_i$ are defined as:
\begin{align} \label{eq:G_a}
\T{(G_1)}{i}{j} &= \T{\Zd}{i}{k} \left( \dbyd{\Zd} \right)^{\!\!k}_{j} & \T{(G_2)}{i}{j} &= \T{Z}{i}{k} \left( \dbyd{Z} \right)^{\!\!k}_{j}  \nonumber \\
\T{(G_3)}{i}{j} &= -\T{\Zd}{k}{j} \left( \dbyd{\Zd} \right)^{\!\!i}_{k} & \T{(G_4)}{i}{j} &= -\T{Z}{k}{j} \left( \dbyd{Z} \right)^{\!\!i}_{k} 
\end{align}
and we use the usual convention for matrix indices given in (\ref{eq:matrixderiv}).
Note that in $G_1$ and $G_2$ the ordering of indices is that of usual matrix multiplication, while for $G_3$ and $G_4$ the opposite is the case. The $G_i$ correspond respectively to each of the terms in (\ref{eq:gauss1}).
The operator $G$ is the infinitesimal generator of the adjoint action
\be
Z \to U Z \Ud , \qquad \Zd \to U \Zd \Ud
\ee
and invariance under this action restricts gauge invariant 
operators to be products of traces of the matrices
 $Z$ and $\Zd$. \newline

\subsection{Geometry of $\cM_2$: coordinates}

In this section and in Section \ref{casfuns} we perform explicit calculations at $N=2$. The motivation for considering small values of $N$ is to perform explicit calculations which shed light on the harder (and more interesting) task of obtaining results at arbitrary finite $N$, a task we return to in Section \ref{genNFacts}.

We start from the Schur decomposition as discussed in Section \ref{sec:schur}, 
\be 
Z ~=~ U  T U^{\dag}  ~=~  L  \tilde T L^{\dagger}.
\ee
In the $N=2$ case $ U(2)/U(1) \cong SU(2) / \mathbb{Z}_2 \cong SO(3) $. We can specify explicit coordinates 
\bea
U &=& \begin{pmatrix} \cos \frac{\th}{2} \, e^{\frac{i}{2}(\phi + \psi)}  &  \sin \frac{\th}{2} \, e^{\frac{i}{2}(\phi - \psi)} \cr 
                     -\sin \frac{\th}{2} \, e^{-\frac{i}{2}(\phi - \psi)} &  \cos \frac{\th}{2} \, e^{-\frac{i}{2}(\phi + \psi)}
\end{pmatrix}  \label{eq:coords1} \\
T &=& \begin{pmatrix}  z_1  &  t_0   \cr 
                        0   &  z_2
\end{pmatrix} . \label{eq:coords2}
\eea
The angles $\th , \f , \psi $ are the  Euler angles  of  $ SU(2)/\mathbb{Z}_2 \cong SO(3)$. With these coordinates $L$ and $\tilde T$ take the form
\bea
L &=& \begin{pmatrix} \cos \frac{\th}{2} \, e^{\frac{i}{2}\phi}  &  \sin \frac{\th}{2} \, e^{\frac{i}{2}\phi} \cr 
                     -\sin \frac{\th}{2} \, e^{-\frac{i}{2}\phi} &  \cos \frac{\th}{2} \, e^{-\frac{i}{2}\phi}
\end{pmatrix}  \label{eq:coords3} \\
\tilde T &=& \begin{pmatrix}  z_1  &  t_0 e^{ i \psi }   \cr 
                        0   &  z_2
\end{pmatrix}. \label{eq:coords4}
\eea
The ranges of the coordinates are 
\begin{align}
&z_1, \, z_2 \in \mathbb{C}, \qquad 0 \le t_0 < \infty, \\
& 0 \le \th \le \pi, \qquad 0 \le \f < 2\pi, \qquad  0 \le \psi < 2\pi .
\end{align}
The Jacobian for the change of variables from $Z_{ij}$ to those above is
\be
J = |z_1 - z_2|^2  \, t_0 \, \sin \th
\ee
and so we have 
\be
\int \prod_{i,j}{ dZ_{ij} d\bar{Z}_{ij} } = \int dz_1 dz_2 dt_0 t_0 dU \, |z_1 - z_2|^2  . \, 
\ee
Note the factor of $t_0$ here which is analogous to the $\int r dr$ one gets when using plane polar coordinates. Here $dU$ is the Haar measure on $SU(2)$ which we integrate out and normalise to 1 in the definition of the measure.
The implication of the measure is that the region $ t_0 =0 $, where 
the orbit structure changes compared to that at $ t_0 \ne 0$, has measure zero.  
Likewise the collision of points $z_1 = z_2$ in 
$ Sym^N ( \mC ) $ has measure zero. 

The invariant line element on $gl(2,\mC)$ is given by
\be 
ds^2 = \tr dZ d\Zd.
\ee
We introduce the notation
\be
\omega = U^{-1} dU = \begin{pmatrix}      \o_{11}    &   \o_{12}  \cr 
                    -\bar{\o}_{12}  &  - \o_{11}
   \end{pmatrix} ,
\ee
and using $\omega^{\dag} = -\omega $ we expand $dZ = U\left( dT + [\o , T] \right) \Ud$.

The line element is then expressible as
\bea 
ds^2 &=& \tr \big( dT + [\o,T] \big) \big( d\Td + [\o,\Td] \big) \\
     &=& \left|dz_1 + t_0 \bar{\o}_{12} \right|^2 + \left|dz_2 - t_0 \bar{\o}_{12} \right|^2 \nonumber \\
     & & + \, \left|dt_0 + 2 t_0 \o_{11} - (z_1 - z_2) \o_{12} \right|^2 + \left|(z_1 - z_2) \o_{12} \right|^2. \label{eq:metriczt}
\eea
Using the Cartan one-forms $\o_i$ on $SU(2)$ (see e.g.~\cite{Ortin:2007CartanSU(2)}),
\be
\o  = U^{-1} dU = - \o_i T_i , \qquad T_j = \frac{i}{2} \s_j,
\ee
one may read off the metric on the orbit; we shall do this in the next section.

As an aside, we note that $U_{12}$ defined below \ref{T2T1} is not a standard permutation matrix in $U(N)$ 
(the reader may check that the standard permutation matrices in $U(N)$ do not
preserve the triangular form).
For concreteness we now exhibit this at $N=2$. Consider the two matrices
\bea
T_1 &=& \begin{pmatrix}  z_1  &  t_0  \cr 
                        0   &  z_2
\end{pmatrix}  \\
T_2 &=& \begin{pmatrix}  z_2  &  t_0  \cr 
                          0   &  z_1
\end{pmatrix}
\eea
where we have chosen $t_0 \in \mathbb{R}$.

Defining $D = \sqrt{t_0^2 + |z_1 - z_2|^2}$, we then have $T_2 = U_{12} T_1 \Ud_{12}$ with 
\be  \label{eq:U12}
U_{12} = \frac{1}{D} \begin{pmatrix} t_0 & -(\bar z_1 - \bar z_2) \cr 
                z_1 - z_2 & t_0 \end{pmatrix} .
\ee
Clearly this is not the standard permutation matrix 
 $ \left(  \begin{smallmatrix} 0&1\\ 1& 0 \end{smallmatrix} \right ) $, 
but it   performs the 
 permutation transformation $z_1 \leftrightarrow z_2$ while preserving the 
 triangular structure. For  $N>2$ 
the analogous transformation does not 
 just permute the $z_i$ entries but transforms the $t_{ij}$ nontrivially.

\subsection{Differential Gauss's law and orbits at $N=2$} 

Using a change of variables, one may express the Gauss Law 
operator $G$ (\ref{Gauss}-\ref{eq:G_a}) in the coordinates defined in (\ref{eq:coords1}-\ref{eq:coords2}). This results in the following
 form of the Gauss's Law operator:
\be \label{eq:diffGauss}
G = \left( \begin{array}{cc} 
-i \dbyd{\phi}       & i e^{i \p} (-\dbyd{\theta} - i \cot \theta \dbyd{\f}  + i \csc \theta \dbyd{\psi} ) \\
i e^{-i \p} (\dbyd{\theta} - i \cot \theta \dbyd{\f}  + i \csc \theta \dbyd{\psi} ) &  i \dbyd{\phi}    \end{array} \right) .
\ee
This must vanish on gauge invariant wavefunctions, which must therefore be functions only of $z_i, t_0$. We will show in Section \ref{truncN=2} that the ring of gauge invariant polynomials has five generators.   

The Gauss's Law reduces the 8D space $gl (2,\mC)$ to the 5D space parametrized by 
$ ( z_1 , z_2 , t_0 ) $. We shall find it convenient to define 
\be \label{eq:zc-z}
z_{c} = z_1 + z_2, \qquad z = z_1 - z_2 .
\ee
As we have seen, we can exchange $z_1, z_2 $ while 
leaving $t_0$ invariant; this means mapping $ z \rightarrow -z $, and so the space of inequivalent orbits is 
\bea 
 \cM_2 = \mC \times ( \mC/ \mathbb{Z}_2 )  \times \mR^+ .
\eea

From the metric (\ref{eq:metriczt}) expressed in terms of $z_i, t_0$ we see that the nature of the orbits changes as we 
move in the space $( \mC/ \mathbb{Z}_2 )  \times \mR^+$. The centre of mass coordinate $z_{c}$ does not affect the nature of the orbits and so we restrict our attention to a $\mathbb{Z}_2$ quotient of the $z, t_0$ space. Let us define
\bea 
X &=& ( \mC/ \mathbb{Z}_2 )  \times \mR^+ \cr 
  &=& X_0 \cup X_1 \cup X_2 \cup X_3 
\eea 
where $X $ is the region in $ (z,t_0)  $ space where $t_0 \geq 0 $, $ Re(z) \geq 0 $, and the subregions $X_i$ are defined as follows:
\begin{itemize}
\item $X_0$ is the subregion $ t_0 > 0, z \ne 0 $
\item $X_1$ is the subregion $ t_0 > 0, z = 0 $
\item $X_2$ is the subregion $ t_0 = 0, z \ne 0 $
\item $X_3$ is the point  $ t_0 = 0, z = 0 $.
\end{itemize} 

The metric on the gauge orbit is determined by fixing $ z_i , t_0$
 in (\ref{eq:metriczt}). On $X_0$ and $X_1$ the orbit is topologically 
$SO(3)$; the metric is complicated in general but on $X_1$ it qualitatively
 resembles the round three-sphere metric. On $X_2$ the orbit is a round
 $S^2$, while on $X_3$ the orbit is a point. This completes the global
description of the parameter space and the orbits. Note that on $X_0$
the metric is regular but on $X_1$, $X_2$ and $X_3$, 
the determinant of the metric is zero.

\subsection{The algebra of functions on $\cM_2$}\label{truncN=2} 

The algebra of functions on $ \cM_{N}$ is generated by single 
trace polynomials in $ Z , Z^{\dagger}$.
In the $N \to \infty$ limit any word in the two letters $ Z , Z^{\dagger} $, 
up to cyclic permutations, corresponds to a single-trace gauge-invariant function 
and hence to a function on $\cM_{\infty}$. 
At finite $N$, traces of long words can be expressed 
in terms of products of traces of shorter words and so the ring of gauge 
invariant functions has a finite set of generators. 

In \cite{Kimura:2007wy} this truncation of the 
generators was discussed in terms of 
degenerations of Brauer algebra projectors. 
Here we investigate these finite $N$ truncations in detail at $N=2$ 
and find that it suffices to apply the
Cayley-Hamilton theorem to obtain the necessary relations.

The Cayley-Hamilton theorem states that a matrix satisfies its own characteristic polynomial. At $N=2$ this means that
\be \label{eq:CH}
Z^2 - (\tr Z) Z + (\det Z)1_2 = 0.
\ee
Taking the trace of this equation gives a relation between $\tr Z^2$, $\tr Z$ and $\det Z$, only two of which are thus algebraically independent as polynomials in the matrix entries. We choose $\tr Z^2$ and $\tr Z$ to be independent, and write
\be
\det Z = \frac{1}{2}\left[ \tr Z \tr Z - \tr Z^2 \right].
\ee
We also have the corresponding equation for $\Zd$. \newline

We claim that the ring of multi-trace GIOs in $Z, \Zd$ at $N=2$ is the polynomial ring generated by the set
\be \label{eq:N=2ring}
\cB = \left\{ \tr Z , \quad \tr Z^2 , \quad \tr \Zd , \quad \tr \Zd{}^2 , \quad \tr Z \Zd  \right\}.
\ee 
In order to prove this, it is enough to show that all other \textit{single trace} operators are algebraically dependent on the operators above.

We prove this in an inductive fashion. Let $W$ to denote any matrix word made from $Z$ and $\Zd$, e.g.~$W=ZZ\Zd Z$. 
Multiply (\ref{eq:CH}) by $W$ and take the trace. This yields the relation
\bea \label{eq:CH2}
\tr (Z^2 W) - (\tr Z) \tr (Z W)  + \frac{1}{2}\left[ \tr Z \tr Z - \tr Z^2 \right] \tr W = 0 .
\eea
This shows that $\tr (Z^2 W)$ is algebraically dependent on $\tr (Z W)$, $\tr W$ and the operators in $\cB$, and similarly, $\tr (\Zd{}^2 W)$ is algebraically dependent on $\tr (\Zd W)$, $\tr W$ and the operators in $\cB$. 

Replacing $Z$ by $Z \Zd$ in (\ref{eq:CH}) and using $\det Z \Zd = \det Z \det \Zd$ gives
\be
\tr (Z \Zd)^2 = (\tr Z \Zd)^2 - \frac{1}{2}\left[ \tr Z \tr Z - \tr Z^2 \right] \left[ \tr \Zd \tr \Zd - \tr \Zd{}^2 \right].
\ee
This shows us that $\tr (Z \Zd)^2$ is algebraically dependent on the operators in the set $\cB$. Similarly, for any word $W_2$ of length at least two, $\tr W_2^2$ is algebraically dependent on $\tr W_2$ and the operators in the set $\cB$.

We conclude that a single trace operator consisting of the trace of a word made from $Z$ and $\Zd$ is algebraically dependent on single trace operators of shorter length iff it contains one of the following combinations as part of the word: 
\be \label{eq:dependent}
Z^2 W , \qquad \Zd{}^2 W , \quad \mathrm{or} \quad W_2^2
\ee
where as above $W$ stands for any (non-zero length) word in $Z$ and $\Zd$, and $W_2$ stands for such a word of length at least two. 

Iterating the above results, a single trace operator containing one of the combinations in (\ref{eq:dependent}) can be expressed as sums of products of shorter and shorter single trace operators until it is expressed as a sum of products of single trace operators containing none of the combinations in (\ref{eq:dependent}).  A maximal set of algebraically independent operators is therefore given by those single trace operators which do not contain any of the expressions in (\ref{eq:dependent}). As claimed this is the set $\cB$.

It is worth remarking that we start with a description of the space $gl(2,\mC)$ 
in terms of polynomials in $z_1 , z_2, t_0, \th, \f, \psi $. 
The differential Gauss Law (\ref{eq:diffGauss}) removes the angular variables leaving the ring of polynomials in the remaining variables, which we denote
\be\label{ringbig} 
\langle z_1 , z_2 , \bar z_1 , \bar z_2,  t_0   \rangle.
\ee
Invariance under large gauge transformations reduces the ring of gauge invariant polynomials to the polynomial ring generated by $\cB$. Recalling the definitions $z_c = z_1 + z_2,~z = z_1 - z_2$ and defining 
\be
\cZ = z^2 , \quad \bar \cZ = \bar{z}^2,  \quad T_0 = t_0^2 + \frac{z\bar z}{2},
\ee
the ring of gauge invariant polynomials is equivalently the polynomial ring 
\be \label{eq:ringT0}
\langle z_{c} ,\bar z_{c} , \cZ , \bar \cZ , T_0 \rangle .
\ee
This is analogous to $U(N)$ gauged Hermitian matrix quantum mechanics
 where the differential Gauss Law reduces to polynomials in the eigenvalues
\be
\langle x_1 , x_2, \ldots , x_N \rangle 
\ee
and invariance under the $S_N$ residual Weyl transformations 
reduces the gauge invariant polynomials to symmetric polynomials in $x_1 , 
x_2,  \cdots , x_N$, equivalently polynomials in the variables
\be
\langle ( x_1 +  x_2 + \cdots + x_N ) , ( x_1^2 + x_2^2 + \cdots + x_N^2 ) ,
 \ldots , (x_1^N + x_2^N + \cdots + x_N^N ) \rangle .
\ee
In the hermitian case, we are going from a ring to a sub-ring, which
 corresponds to going from the  space  $\mR^N$ to its 
quotient space  $ \mR^N / S_N$. In our model, we are going from the ring
(\ref{ringbig}) to the sub-ring (\ref{eq:ringT0}), and correspondingly from the 
$\mR^4 \times \mR^+ = \mC^2 \times \mR^+  $ parametrized by the five coordinates $ z_i , t_0$ to 
$\cM_2 $.   Because of the off-diagonal degrees of freedom,  
$\cM_2$ is not a  straightforward quotient of $\mR^4 \times \mR^+$.

A full investigation of finite $N$ relations for  $N >2$ 
is left for the future. We expect it will be useful to 
combine the Cayley-Hamilton approach with the 
the vanishing of the Brauer projectors, such as in 
equation (8.16) of \cite{Kimura:2007wy}.

\section{Free particle structures and counting on $\cM_2$} 

\label{casfuns} 

The remainder of the paper involves the Brauer basis for complex matrix models constructed in \cite{Kimura:2007wy}.
A Brauer basis operator is a linear combination of multi-trace operators;
it is a polynomial of degree $m$ in $Z$ and degree $n$ in $\Zd$. A brief 
review of the essential properties of the basis and some simple examples are 
given in Appendix \ref{app:Brauer}, where references to existing literature are
also given. 

A Brauer basis operator is written as 
\be \label{eq:BrauerOp}
\cO^{\g}_{\a,\b;i,j} (Z,\Zd)
\ee
where $\a$ and $\b$ are Young diagrams with $m$ and $n$ boxes and $\g = (k, \g_+, \g_-)$ where $k$ is an integer in the range $0 \le k \le \min(m,n)$ and $\g_+,\g_-$ are Young diagrams with $m-k$ and $n-k$ boxes respectively. For a more complete explanation of the labels please see Appendix \ref{app:Brauer}.

As an example for the reader to bear in mind, when $(m,n) = (1,1)$, suppressing non-essential labels the Brauer basis is
\bea
\cO^{k=0}_{[1],[\bar{1}]} (Z,\Zd) &=& \tr Z \tr \Zd - \frac{1}{N} \tr Z \Zd \\
\cO^{k=1}_{[1],[\bar{1}]} (Z,\Zd) &=& \frac{1}{N} \tr Z \Zd  ~.
\eea
Since we discuss in particular the label $k$ throughout the rest of this paper, the following comment will be useful to the uninitiated reader.
In the construction of the Brauer basis, a term with a single `$Z\Zd$'
inside the same trace, such as $\tr Z Z\Zd$, involves a single `Brauer contraction'. 
Terms such as $\tr Z\Zd \tr Z\Zd \Zd$ or $\tr Z\Zd  Z Z\Zd$ involve two such Brauer contractions, etc.

The label $k$ is related to the number of contractions as follows. 
If one writes a Brauer basis operator as a sum 
of terms in order of increasing contractions, as the two operators
above are written, 
an operator with label $k$ begins with a term involving $k$ Brauer contractions.
 We have not proved this, but we believe it to be true from all the 
examples we know. 
Thus
 the leading term in a $k=0$ operator is the product of a purely holomorphic operator and a purely anti-holomorphic operator, while all terms in a $m=n=k$ operator involve $k$ contractions.

The first result of this section is a conjecture for the 
solution to the $N=2$ counting of the operators of the Brauer 
basis, for which which we provide numerical evidence. 

The second result will be to find evidence of a `free fermions on a circle'
structure in the $k=0$ sector. This generalizes to 
any $N$, as discussed in Section \ref{sec:k=0generalN}. 
We show that the Brauer basis at $N=2$ 
can be neatly expressed in terms of five integers and observe the correspondence between 
states in the $k=0$ sector and two free fermions on a circle.
In these developments 
a crucial role is played by the structure of the ring of Casimirs.

The third result in this section is to show that the momenta of the free fermions of the $k=0$ sector 
can be constructed from differential operators 
in variables which include both eigenvalues and off-diagonal elements of $Z$. 
This leads us to observe that the complex matrix model 
contains free fermions arising in a novel way, different 
from the way they arise in hermitian or unitary models.

\subsection{Casimir operators and a ring of degree-preserving differential 
operators}\label{finitesetgens}  

The differential operators introduced in equation (\ref{eq:G_a})
were studied  in \cite{Kimura:2008ac} as generalized Casimirs commuting 
with the scaling operator for $ Z , Z^{\dagger}$,
which is the  Hamiltonian for zero coupling SYM. 
This ring is analogous to the ring generated
 by $\cB$ in Section \ref{truncN=2}; at $N=2$ the generating set is 
\be \label{eq:DPGIDOs}
\cD =  \left\{ \tr G_2 , \quad \tr G_2^2 , \quad \tr G_3 ,
\quad \tr G_3^2 , \quad \tr G_2 G_3  \right\}
\ee
where  $G_2, G_3$ were defined in (\ref{eq:G_a}) 
\be
\T{(G_2)}{i}{j} = \T{Z}{i}{k} \left( \dbyd{Z} \right)^{\!\!k}_{j} \qquad\qquad \T{(G_3)}{i}{j} = -\T{\Zd}{k}{j} \left( \dbyd{\Zd} \right)^{\!\!i}_{k}.
\ee
Defining 
\be
G_L = G_2+ G_3 ~,
\ee
we introduce the Hamiltonians
\begin{align} \label{eq:Hamiltonians}
H_1       &= \tr G_2  &       H_2 &= \tr G_2^2  \nonumber \\
\bar{H}_1 &= \tr G_3  & \bar{H}_2 &= \tr G_3^2 & H_L &= \tr G_L^2 ~.
\end{align} 
Each of these operators commutes with the scaling operator for $Z$ and $ Z^{\dagger}$,
which is $ H = H_1 + \bar H_1$. The operators in $\cD$  generate a ring of 
commuting Hamiltonians related to the integrability of the system. 
We have defined $H_L$ for later convenience; its name derives from the fact that the operator $G_2+ G_3$ is the infinitesimal generator of the left action of $U(N)$ \cite{Kimura:2008ac}:
\be
Z \rightarrow UZ, \qquad \Zd \rightarrow \Zd \Ud.
\ee
It was shown in \cite{Kimura:2008ac} that the five
 operators defined in (\ref{eq:Hamiltonians}),
\be \label{eq:CasimirOps}
\cH_A = \Big\{ 
H_1, \quad  \bar{H}_1, \quad H_2 , \quad  \bar{H}_2 , \quad  H_L \Big\} 
\ee
measure respectively the Casimirs 
\be \label{eq:Casimirs}
\cC_A = \Big\{ C_1(\a),  \quad  C_1(\b),  \quad  C_2(\a),  \quad  C_2(\b), \quad  C_2(\g) \Big\} ~.
\ee
Generalized Casimir operators such as $\tr (G_{2}^{2}G_{3})$ were investigated in \cite{Kimura:2008ac} and were shown to be sensitive to the labels $i,j$ in (\ref{eq:BrauerOp}).
Since the matrix elements of $G_2$ and $G_3$ commute,
 we may regard $G_2$ and $G_3$ as matrices of c-numbers and apply the Cayley-Hamilton theorem
 as in Section \ref{truncN=2} to show that the set $\cD$ is a maximal algebraically independent set of degree-preserving gauge invariant differential operators.

This observation implies that the generalized Casimir operators such as $\tr G_2^2G_3$  do not yield independent information about the wavefunctions at $N=2$, i.e.~that all the information in the labels $\{\a,\b,\g,i,j\}$ is in fact contained only in $\{\a,\b,\g\}$. We can interpret this fact in terms of Brauer algebra representation theory as follows.

In the restriction of an irreducible representation $\g$ of the Brauer algebra to the representation $A = (\a,\b)$ of $\mC[S_m \times S_n]$, there enters an integer multiplicity $M^{ \g ;  N}_{A}$ defined by
\be
V_{\gamma}^{B_{N}(m,n)}= \bigoplus_{A} M^{\g ; N }_A \, V_{A}^{\mathbb C(S_{m}\times S_{n})}.
\ee
For large $N$, i.e.~$m+n < N$, we denote this multiplicity by $M^{\g}_A$ or $M^{\g}_{\a,\b}$ and using  $\delta \vdash k$ to denote that $\delta$ is a partition of $k$ we have the formula \cite{Koike:1989}
\be \label{eq:multgg}
M^{\g}_A ~=~ M^{\g}_{\a,\b} ~=~  \sum_{ \delta \vdash k }  \sum_{ \delta}  g (  \gamma_+ ,  \delta ; \alpha  ) g ( \gamma_-  , \delta  , \beta   )
\ee
where $g (  \gamma_+ ,  \delta ; \alpha  )$ is a Littlewood-Richardson coefficient.

As reviewed in Appendix \ref{app:Brauer} the indices $i,j$ on a Brauer operator range over the values 
$\{1,\ldots,M^{\g ; N }_A\}$, and so 
the redundancy of the $i,j$ labels at $N=2$ 
means that  $M^{\g ; N=2 }_A  $ is either 0 or 1
for all $\g, A$. A direct proof of this by using the 
finite $N$ constraints on the states of the Brauer representation 
in \cite{Benkart:1992} would be interesting to obtain. 
At this point we will take a more pragmatic  perspective,  
assume it is true, and will find that it leads to 
a consistent counting of states of the complex matrix model at 
$N=2$.

\subsection{Counting of states at $N=2$ and Brauer basis labels }\label{countN=2} 

The ring of gauge invariant operators at $N=2$ is generated by five single trace operators (\ref{eq:N=2ring}). Hence the number of linearly independent multi-trace operators $Q_{mt}^{N=2}( m ,n )$ for fixed $(m,n)$ is counted by the generating function 
\bea  \label{eq:N=2genfunction}
{ 1 \over ( 1 - x ) ( 1- y ) ( 1 - x^2 ) ( 1 - y^2 ) ( 1 - xy ) } = \sum_{ m , n } Q_{mt}^{N=2}( m ,n ) x^m y^n  .
\eea 
This is the Plethystic Exponential \cite{Benvenuti:2006qr,Feng:2007ur} of the single trace generating function
\be
\sum_{ m , n } Q_{st}^{N=2}( m ,n ) \, x^m y^n   ~=~  1 + x + y + x^2 + y ^2 + xy
\ee
derived from the independent single traces in the basis $\cB$ (\ref{eq:N=2ring}).

Having found the  $N=2$ counting of multi-traces, we can express it in terms 
of constraints on the large $N$ Brauer counting. The obvious constraint
$ c_1 ( \g_+ ) + c_1 ( \g_- ) \le 2 $ is  not sufficient. We have argued above that the multiplicities
 $ M^{\g ; N=2}_{\a , \b} $ are either $0$ or $1$. 
We first set
\bea
M^{\g ; N=2}_{\a , \b} &=& \begin{cases} ~1 & \qquad \mathrm{if}~ M^{\g}_{\a , \b} > 0 \\
                                         ~0 & \qquad \mathrm{otherwise}
                                         \end{cases}
\eea
where $M^{\g}_{\a , \b}$ is given by (\ref{eq:multgg}). Having done this we also find it necessary to impose extra constraints on the labels $\a, \b$ for agreement with (\ref{eq:N=2genfunction}). 

The constraints on $\a , \b$ are as  follows. Denoting the length of the $p^{th}$ column of a Young diagram $R$ by $c_p(R)$,  we constrain:
\begin{enumerate}
\item $c_1(\a) + c_1(\b)  \leq  N + k$
\item $[c_1(\a) + c_1(\b)] + [c_2(\a) + c_2(\b)] \leq  2N + k$   
\end{enumerate}
\qquad $\vdots$

and in general for each $p = 1,2,\ldots,\min(m,n) $ , constrain
\be \label{eq:N=2newconstraint}
\qquad\qquad \sum_{r=1}^{p}(c_r(\a) + c_r(\b)) \leq  pN + k. \qquad\qquad  
\ee
We have used SAGE and Mathematica to enumerate all possible Brauer basis operators subject to the constraint (\ref{eq:N=2newconstraint}) and to compare with the Trace basis generating function. The two agree up to $(m,n) = (15,15)$ which is the practical limit for a desktop computer. This conjecture generalizes the `Non-chiral Stringy Exclusion Principle' introduced in \cite{Kimura:2007wy}.  This counting of operators at $N=2$ implies 
a result for the reduction multiplicities $M^{\g , N=2 }_A $, namely that
\bea \label{eq:multN=2star}
M^{\g ; N=2}_{\a , \b} &=& \begin{cases} ~1 & \qquad \mathrm{if}~ M^{\g}_{\a , \b} > 0 ~~\mathrm{and~(\ref{eq:N=2newconstraint})~holds}\\
                                         ~0 & \qquad \mathrm{otherwise}
                                         \end{cases}
\eea
We will re-state this result after simplifying the condition 
(\ref{eq:N=2newconstraint}). 

We can use the fact that the Brauer basis diagonalises the five Casimirs $\cH_A$ to explicitly enumerate the independent Brauer operators at $N=2$. Fixing $(m,n)$ we pick a basis of multi-traces. 
Acting  with the explicit form of the Casimirs (\ref{eq:N=2ztCasimirs}), we can  construct linear combinations which 
are eigenstates of the Casimirs.   Because the eigenvalues of the five Casimirs determine the labels $\a, \b, \g$ uniquely, we can also read off the labels of the allowed operators. We have carried out this procedure for selected values of $(m,n)$ up to $(m,n) = (4,3)$.

\subsection{The Brauer basis 
labels at $N=2$ in terms of five integers}\label{Brauer5}

In Section \ref{truncN=2} we described the states of the $ N=2$ theory 
as generated by a finite set of traces. In this section we will obtain 
the description in terms of the Brauer basis for multi-traces. 
For general $N $, we give a review of the Brauer basis states 
in Section \ref{app:Brauer}. For ease of notation we denote
$r_i=r_i ( \alpha )$ and $ \bar r_i =  r_i ( \beta ) $.

We can choose different sets of five integers to parameterise
 the states, such as  
\bea \label{eq:5a}
&& r_1 , r_2 , \bar r_1 , \bar r_2 , r_1^{\g} \\ 
 \label{eq:5b}
&& r_1^{\g}, r_2^{\g}, k,  r_1, \bar r_1 \\
&&  \label{eq:5c}
r_1^{\g}, r_2^{\g}, k,  r_1, \bar r_2 .
\eea
We will show that each of the above sets of five integers
determines a state uniquely, and we will give the 
constraints on the integers.

A state is determined uniquely at $N=2$ by $\a, \b, \g$, 
containing the set of integers 
\be   \label{eq:fullset}
\left\{ r_1, r_2; ~ \bar{r}_1, \bar{r}_2; ~ k, r_1^{\g}, r_2^{\g} \right\}.
\ee
From the Brauer algebra representation theory briefly reviewed 
 in Appendix \ref{app:Brauer}, we have the following relations : 
\begin{align}
& \sum_i r_i             = m  ,      &   &\sum_i \bar{r}_i  = n,        \label{eq:m,n}         \\
& \sum_i r_i (\g_+)   = m-k   ,  &   &\sum_i r_i (\g_-)   = n-k.     \label{eq:m-k,n-k}    
\end{align}
Using the relationship between $r_i (\g)$, $r_i (\g_+)$ and $r_i (\g_-)$ we have
\be \label{eq:m-n}
\sum_i r_i (\g) ~~=~~ \sum_i r_i (\g_+) - \sum_i r_i (\g_-) ~~=~~ m-n
\ee
which at $N=2$  reads
\be \label{eq:m-nN2}
r_1^{\g} + r_2^{\g} ~=~ m-n.
\ee
Adding the two expressions in (\ref{eq:m-k,n-k}) we find  that 
\be
\sum_i | r_i (\g) |  ~=~ \sum_i r_i (\g_+) + \sum_i r_i (\g_-) ~=~ m+n - 2k
\ee
which at $N=2$ gives 
\be \label{eq:k-rigamma}
k = \frac{1}{2}\left( m +n  - |r_1^{\g}| - |r_2^{ \g} | \right).
\ee
We now show that each of (\ref{eq:5a})-(\ref{eq:5c}) are enough to determine the state via (\ref{eq:fullset}):
\begin{enumerate}
\item Starting from the five integers in (\ref{eq:5a}), we deduce $m,n$ from (\ref{eq:m,n}), $r_2^{\g}$ from (\ref{eq:m-nN2}) and $k$ from (\ref{eq:k-rigamma}).
\item Starting from (\ref{eq:5b}) we read off $r_i(\g_+)$ and $r_i(\g_-)$ by inspecting whether $r_1^{\g}$ and $r_1^{\g}$ are positive or negative. We then deduce $m$ and $n$ from (\ref{eq:m-k,n-k}) and $r_2$ and $\bar{r}_2$ from (\ref{eq:m,n}).
\item Starting from (\ref{eq:5c}) we proceed as in point 2 above.
\end{enumerate}
This shows that each of the three sets of five integers identified are sufficient to identify any state.

\subsubsection{$N=2$ constraints in terms of five integers}

Let us consider the case where $k$ is one of our five integers. We rewrite the $N=2$ constraint (\ref{eq:N=2newconstraint}) as a lower bound on $k$:
\be 
k \ge \sum_{r=1}^{p}(c_r(\a) + c_r(\b)) - 2p  \qquad \mathrm{for~each~} p = 1,\ldots \min(m,n).
\ee
Note that as $p$ increases the lower bound on $k$ gets stronger only when 
\be
c_p(\a) + c_p(\b) > 2.
\ee
Before presenting a general expression for the lower bound on $k$ we examine in detail the case 
\be
0 < r_2 < \bar{r}_2 < r_1 < \bar{r}_1.
\ee
We observe that
\begin{itemize}
\item For $1 \le p \le r_2$ we have $c_p(\a) + c_p(\b) = 4$
\item For $r_2 < p \le \bar{r}_2$ we have $c_p(\a) + c_p(\b) = 3$
\item For $p > \bar{r}_2$ we have $c_p(\a) + c_p(\b) \le 2$
\end{itemize}
The strongest lower bound on $k$ is therefore at $p = \bar{r}_2$ where we have
\bea
k & \ge &  4  r_2 + 3 ( \bar{r}_2 - r_2 ) - 2 \bar{r}_2 \nn \\
\Rightarrow k & \ge & r_2 + \bar{r}_2 .
\eea
Proceeding similarly we find a general expression for the lower bound on $k$.  For simplicity, wlog suppose $r_2 \le \bar{r}_2$. There are three cases to consider:
\begin{enumerate}
\item $r_2 \le r_1 \le \bar{r}_2 \le \bar{r}_1  \qquad  \Rightarrow  \qquad  k \ge r_1 + r_2 $
\item $r_2 \le \bar{r}_2 \le r_1 \le \bar{r}_1  \qquad  \Rightarrow  \qquad  k \ge r_2 + \bar{r}_2 $
\item $r_2 \le \bar{r}_2 \le \bar{r}_1 \le r_1  \qquad  \Rightarrow  \qquad  k \ge r_2 + \bar{r}_2 $.
\end{enumerate}
Combining these we obtain the lower bound
\bea \label{eq:klowerboundnew}
\fbox{$ \quad   
k \ge \min ( r_2 , \bar r_2 ) + \min \left( ~ \min(r_1, \bar{r}_1) , \, \max ( r_2 , \bar r_2 ) ~ \right)   \quad 
$
}
\eea
which is equivalent to  (\ref{eq:N=2newconstraint}). 
We can also express the constraint (\ref{eq:klowerboundnew}) in terms of the five integers in (\ref{eq:5a}) by substituting for $k$ from (\ref{eq:k-rigamma}) to find
\be
\frac{1}{2}\left( m +n  - |r_1^{\g}| - |m-n-r_1^{ \g} | \right) \ge \min ( r_2 , \bar r_2 ) + \min \left( ~ \min(r_1, \bar{r}_1) ,  \max ( r_2 , \bar r_2 ) ~ \right) .
\ee
We can now re-state the result (\ref{eq:multN=2star}) for the $N=2$ reduction multiplicities: 
\bea
 \fbox{$ \quad
M^{\g ; N=2}_{\a , \b} = \begin{cases} ~1&  \qquad \mathrm{if}~ M^{\g}_{\a , \b} > 0 ~~\mathrm{and~(\ref{eq:klowerboundnew})~holds \quad }  \\
                                         ~0 & \qquad \mathrm{otherwise.}
                                         \end{cases}
$}
\eea

\subsection{The Casimirs as differential operators in  $z_i , t_0 $}\label{casdiffzt} 
	
In this section we express the Casimir operators / Hamiltonians from Section {\ref{finitesetgens}} as differential 
operators on $ \cM_2$. 

Below are calculated expressions in the coordinates $z_i, t_0$ for the Hamiltonians defined in (\ref{eq:Hamiltonians}).
For convenience define 
\begin{align} \label{eq:L1L2Lt}
 L_1 &= z_1 \dbyd{z_1}  & \Lb_1 &= \zb_1 \dbyd{\zb_1} &   \\
 L_2 &= z_2 \dbyd{z_2}  & \Lb_2 &= \zb_2 \dbyd{\zb_2} &  L_t &= \frac{t_0}{2} \dbyd{t_0}    
\end{align}
and recall the notation $z_c = z_1 + z_2 , ~z = z_1 - z_2$.

Recalling the definition $G_L = G_2+G_3$ from above (\ref{eq:Hamiltonians}), we find the following expressions:
\bea
H_1 ~=~ ~~~ \tr G_2 &=& L_1 + L_2 + L_t   \\ \cr
\bar H_1 ~=~ - \tr G_3 &=&  \bar{L}_1 + \bar{L}_2 + \bar{L}_t  
\eea
\bea
H_2 ~=~ \tr G_2^2 &=& L_1^2 + L_2^2 + \left( 1 - \frac{2 z_1 z_2 \zb }{z t_0^2} \right) L_t^2 \nonumber \\ 
             &+& \frac{2}{z} \left( z_1 L_1 - z_2 L_2 \right) L_t + \frac{z_c}{z} \left( L_1 - L_2 \right) + L_t \\
& & \nonumber \\
H_3 ~=~ \tr G_3^2 &=& \overline{ \tr ( G_2^2 ) } \\
& & \nonumber \\
H_L ~=~ \tr G_L^2 &=& 
(L_1 - {\Lb_1} )^2 + (L_2 - {\Lb_2} )^2 
+ \frac{z_c}{z} \left( L_1 - L_2 \right) +
 \frac{\zb_c}{\zb} \left( {\Lb_1} - {\Lb_2} \right) \nn \\
                        &-& \frac{2}{|z|^2} 
\biggl \{  ~~  
 t_0^2 (L_1 - L_2)(\Lb_1 - \Lb_2) + 
\frac{1}{t_0^2} (z_1 \zb_1 - z_2 \zb_2)^2 L_t^2  \nn   \\ 
&& - (z_1 \zb_1 - z_2 \zb_2 )
 \left[ (L_1 - L_2) + (\Lb_1 - \Lb_2 ) \right] L_t    
 - (z_1 \zb_1 + z_2 \zb_2) L_t  ~~ \biggr \} \nn \\    
  \label{eq:N=2ztCasimirs}
\eea
Some useful formulae in doing these calculations are now given.
Recall from (\ref{dZLtT}) the definition $ V  = L^{\dagger} dL $ and the expression
\bea 
dZ &=& L \left( d\tilde T + [ V , \tilde T ] \right) L^{\dagger}  .
\eea 
Defining 
\be
d \tilde X  =  d \tilde T + [ V , \tilde T ] \qquad \mathrm{and} \qquad \T{ (\tilde G_2) }{i}{j} = \T{\tilde T}{i}{p} \left( { \d \over \d  \tilde X } \right)^{\!\!p}_j
\ee
one may derive
\bea 
\T{dZ}{i}{j} &=& \T{L}{i}{p}  \T{d\tilde{X}}{p}{q}   \T{L^{\dagger}}{q}{j}    \cr 
\left( { \d \over \d Z}  \right)^{\!\!i}_j  &=& \T{L}{i}{p} \T{L^{\dagger}}{q}{j} \left( \dbyd{\tilde X} \right)^{\!\!p}_q  \cr
\T{(G_2)}{i}{j}  &=& \T{L}{i}{p} \T{L}{q}{j}   \T{(\tilde G_2 )}{p}{q}.
\eea 
The computation of $\T{(\tilde G_2 )}{p}{q} $ shows that
it contains angular derivatives. When we calculate 
\bea
\tr G_2^2 &=&  \T{L}{i}{p} \T{L}{q}{j} \T{(\tilde G_2 )}{p}{q}  \T{L}{i}{r} \T{L}{s}{j} \T{(\tilde G_2 )}{r}{s}
\eea
it is important not to neglect the terms obtained from the action of these angular derivatives from $\T{(\tilde G_2 )}{p}{q} $
 on $ \T{L}{i}{r} \T{L}{s}{j} $. 

\subsubsection{The Casimirs  as operators on polynomial rings} 

We observed in equation (\ref{eq:ringT0}) that the multi-trace operators built from $Z , Z^{\dagger} $ form a polynomial ring whose generators we may take to be
\be
z_c , \quad \bar z_c , \quad \cZ = z^2 , \quad \bar \cZ = \bar{z}^2,  \quad T_0 = t_0^2 + \frac{z\bar z}{2}.
\ee
The above differential operators $H_2, H_3, H_L$ map polynomials in these variables to polynomials.  Changing  variables to these generators makes this 
manifest: 
\bea 
H_2 &=& 2 L \left( L + { 1 \over 2 } \right) + { 1 \over 2 } L_c ( L_c + 3 ) 
 + L_0 ( L_0 +1 ) + { 2 z_c^2 \over \cZ } L + { z_c^2 \over \cZ } ( 2 L -1 ) L \cr 
&&+ { \cZ \over 2 z_c^2 } L_c ( L_c -1 ) 
+ {  \bar \cZ \over 8 T_0^2 } ( z_c^2 - \cZ ) L_0 ( L_0 -1 ) 
 + 2 \left( 1 + { z_c^2 \over \cZ } \right) L L_0 + 2 L_0 L_c + 4 L L_c . \nn\\ 
\eea 
where $ L = \cZ { \d \over \d \cZ } $, 
$ L_0 = T_0 { \d \over \d T_0 } $ and $ L_c = z_c { \d \over \d z_c } $. 
$H_3$ is obtained by complex conjugation and the same exercise can also be done for $H_L$
to illustrate that they are operators that map polynomials to polynomials.

\subsection{Eigenvalues of the Casimir operators}\label{eigcasN=2}  

As reviewed in Section \ref{sec:review}, a Young diagram $R$ with non-negative row lengths $r_i$ labels energies $\cE_i$ of $N$ fermions in a one-dimensional harmonic oscillator potential, given by
\be \label{eq:energies-SHO2}
\cE_i  =  r_i  +  ( N - i )
\ee
and a Young diagram ($N$-staircase) $R$ with arbitrary integer $r_i$ labels momenta $p_i$ of $N$ free fermions on a circle 
given in terms of the Fermi energy $n_F = \frac{N-1}{2}$ by
\be \label{eq:energies-circle2}
p_i  =  r_i  +  ( n_F + 1 - i ) ~.
\ee

In this section we review the fact that the values of 
\begin{itemize}
\item the $N$ independent $U(N)$ Casimirs $C_i(R)$ of the representation $R$ 
\item the $N$ row lengths $r_i$, and
\item the $N$ corresponding fermion momenta $p_i$  are equivalent data.
\end{itemize}
The same remark holds for non-negative $r_i$ with $p_i$ replaced by $\cE_i$.

In Section \ref{finitesetgens} we introduced differential operators studied in \cite{Kimura:2008ac} which when acting on a Brauer basis function $\cO^{\g}_{\a\b}(Z,\Zd)$ measure the quadratic Casimir of the Young diagrams $\a, \b, \g$. Given a $U(N)$ Young diagram $R$, its linear and quadratic Casimirs are
\bea
C_1(R) &=& \sum_i{ r_i }  \,\,\, =  \,\,\,  n    \label{eq:c1}  \\
C_2(R) &=& nN + \sum_i{ r_i (r_i-2i+1) }.   \label{eq:c2}
\eea

Using the definition of $p_i$ (\ref{eq:energies-circle2}) we can write $C_2$ as
\bea \label{eq:C2-circle}
C_2(R) &=& \sum_{i=1}^{N} p_i^2 - \frac{N}{12}(N^2-1)
\eea
which agrees with \eq{eq:H_f-UMM}. Using the definition of $\cE_i$ (\ref{eq:energies-SHO2}) we can also write $C_2$ as
\bea \label{eq:C2-SHO}
C_2(R) &=& \sum_{i=1}^{N} \cE_i^2 - (N-1) n - \frac{N}{6}(N-1)(2N-1) ~.
\eea
For general $N$, knowledge of the values of the $N$ independent Casimir invariants $C_i$ determine the values of the
power sum symmetric polynomials 
\be \label{eq:pN}
\cP_a = p_1^a + p_2^a + ..... + p_N^a 
\ee
which in turn for $a = 1,\ldots,N$ enables us to solve for $p_i$ or respectively $\cE_i$ (see e.g.~\cite{Zelobenko:1973}). \newline

We now demonstrate this in the $N=2$ theory. The free fermions on a circle have ground state with energy $p_1 = \ha, \,\, p_2 = -\ha$ and in general we have 
\be \label{eq:energies-circleN=2}
p_1  =  r_1 + \ha,    \qquad    p_2 =  r_2 - \ha ~.  
\ee
Setting $N=2$ in (\ref{eq:c2}) gives
\be \label{eq:C2ri}
C_2 = r_1 (r_1 + 1) + r_2 (r_2 - 1)
\ee
and so we may express $C_1$ and $C_2$ in terms of $p_i$ as
\bea
C_1 &=& p_1 + p_2     \nn  \\  
C_2 &=& p_1^2 + p_2^2  - \ha ~.  \label{eq:c2pi}
\eea
The resulting quadratic equations for $p_i$ in terms of $C_1$ and $C_2$ have solution
\bea 
p_1 &=&  \frac{C_1}{2} + \sqrt { {C_2 \over 2 } - { C_1^2 \over 4 } +{ 1 \over 4 } }   \nn \\
p_2 &=&  C_1 - p_1. \label{eq:piCi}
\eea

\subsection{The $k=0$ sector}\label{sec:N=2,k=0}  

In the $k=0$ sector $\g = (0, \a, \b)$ so operators are labelled simply by $\a$ and $\b$ which are representations of $ S_m $ and $S_n$ respectively. 
To connect with the notation of the unitary matrix model,
 we write $\a = R$ and $\b =  S$.
If $ S = \emptyset $, then the $k=0$ operator is the holomorphic Schur polynomial corresponding to the representation $R$:
\be
\cO^{k=0}_{R,\emptyset}(Z,\Zd) = \chi_R(Z)~.
\ee
If $R = \emptyset $, then the $k=0$ operator is the anti-holomorphic Schur polynomial corresponding to the representation $\bar S$:
\be
\cO^{k=0}_{\emptyset,\bar S}(Z,\Zd) = \chi_{ S}(\Zd)
\ee
and if both $\a$ and $\b$ are nontrivial, the leading order term in the expansion of $\cO^{k=0}$ begins with the product of the holomorphic and antiholomorphic Schur polynomials:
\be \label{eq:k=0form}
\cO^{k=0}_{R,\bar S}(Z,\Zd) = \chi_R(Z)\chi_{ S}(\Zd) + \cdots ~,
\ee
where the dots denote terms with at least one $Z \Zd$ inside a trace as discussed at the start of Section \ref{casfuns}.

There is an isomorphism between the $k=0$ sector and the states of the Unitary matrix model \cite{Kimura:2007wy}:
\be\label{k0ops} 
\cO^{k=0}_{R \bar S}(Z, \Zd) \longleftrightarrow  \chi_{R \bar S} (U)
\ee
which is obtained by replacing $Z$ with a unitary matrix:
\be\label{k0ops2} 
\cO^{k=0}_{R \bar S}(U, \Ud) = d_R d_S \chi_{R \bar S} (U) ~.
\ee
The two point functions of both sets of operators are diagonal; up to a choice of normalisation,
\be \label{eq:2ptfunctions}
\Braket{\cO^{\dagger \, k=0}_{R \bar S}(Z, \Zd)}{\cO^{k=0}_{R' \bar S'}(Z, \Zd)} 
 \quad = \quad \Braket{\chi^{\dag}_{R \bar S}(U)} {\chi_{R' \bar{S'}}(U)} \quad = \quad \delta_{R R'} \delta_{\bar{S} \bar{S'}} 
\ee
and the reader familiar with the `coupled characters' studied in two-dimensional Yang-Mills will notice that the structure of \eq{eq:k=0form}
is of the same form as the coupled character $\chi_{R \bar S}$.
The $k=0$ states are thus isomorphic to the states of $N$ free fermions on 
a circle via the map given in the same section.
 
At $N=2$, the label $\g_c$ as defined in \eq{eq:gammac} may have at most two rows, $r_1^{\g},r_2^{\g}$ and so the integers $(k=0, r_1^{\g},r_2^{\g})$ are enough to specify an operator. The list of all $N=2$ operators for given $(m,n)$ in Appendix \ref{app:gammalist} shows that:
\begin{itemize}
\item If $ r_1^{\g} > 0 , r_2^{\g } \ge 0 $, 
then $\beta = \emptyset $ and we have 
a holomorphic Schur polynomial. 
\item If $  r_1^{\g} \le 0 ,   r_2^{\g } < 0 $ then $\a = \emptyset $ and we have 
an antiholomorphic Schur polynomial. 
\item If $  r_1^{\g} > 0 ,   r_2^{\g } < 0 $ then the operator is of the form \eq{eq:k=0form}. At $N=2$ there is a unique such operator.
\end{itemize}
Since row lengths and fermion momenta are equivalent data in specifying a state, 
the above constraints may be rewritten in terms of fermion momenta $p_i^{\g}$.
In the next section, we will see how the momenta of 
these fermions can be expressed in terms of differential operators 
in $z_i, t_0$.

\subsection{Free particle momenta as functions of differential operators} \label{sec:momentadiff}

As noted in (\ref{eq:CasimirOps}), when applied to an $N=2$ Brauer basis operator $\cO^{\g}_{\a,\b}$,  the differential operators 
\be \label{eq:CasimirOps2}
\cH_A = \Big\{ 
H_1, \quad  \bar{H}_1, \quad H_2 , \quad  \bar{H}_2 , \quad  H_L \Big\} 
\ee
measure the values of the Casimirs
\be
\cC_A = \Big\{ C_1(\a),  \quad  C_1(\b),  \quad  C_2(\a),  \quad  C_2(\b), \quad  C_2(\g) \Big\}
\ee
respectively. We also have the fact that $C_1(\g)$ is measured by $H_1 - \bar{H}_1$. We define fermion momentum operators 
\be
\hat{p}_A = \Big\{ \hat{p}_1, \quad \hat{p}_2, \quad \hat{\bar p}_1, \quad \hat{\bar p}_2,\quad  \hat{p}_1^{\g},\quad \hat{p}_2^{\g} \Big\}
\ee
whose eigenvalues are $ p_1, \,\, p_2 , \,\, \bar p_1, \,\, \bar{p}_2, \,\, p_1^{\g},  \,\, p_2^{\g}$ respectively. We now repeatedly apply (\ref{eq:piCi}) to each of $\a,\b,\g$ in turn which enables us to derive expressions 
for these operators in terms of the basic gauge invariant operators $ \cH_A$. 

Applying (\ref{eq:piCi}) to the label $\a$ and promoting to an operator equation we obtain
\bea 
\hat{p}_1 &=&  \frac{H_1}{2} + \sqrt { {H_2 \over 2 } - { H_1^2 \over 4 } +{ 1 \over 4 } } \nn  \\
\hat{p}_2 &=&  H_1 - \hat{p}_1 .  \label{eq:piHi}
\eea 
Applying (\ref{eq:piCi}) to the label $\b$  we obtain analogous expressions for $\hat{\bar p}_1 , \hat{\bar p}_2$ in terms of $\bar H_1 , \bar H_2$. 

Applying (\ref{eq:piCi}) to the label $\g$, promoting to an operator equation and defining $\hat{d} = H_1 - \bar{H}_1$ we obtain  
\bea\label{pi-gamma-diff}
\hat{p}_1^{\g} & =&     \frac{\hat{d}}{2} + \sqrt { {H_L \over 2 } - { \hat{d}^2 \over 4 } +{ 1 \over 4 } } \cr    
\hat{p}_2^{\g} & = &  \hat{d} - \hat{p}_1^{\g}
\eea

As noted in Section \ref{sec:N=2,k=0}, in the $k=0$ sector 
a state is specified simply by the values of the row lengths 
$r_1^{\g}, r_2^{\g}$, or equivalently by
the values of the fermion momenta $p_1^{\g}, p_2^{\g}$ and so 
we now identify $\hat p_1^{\g},\hat p_2^{\g}$
as formal expressions for the
momenta of the $k=0$ fermions on a circle.
We shall extend this result to arbitrary $N$ in the next section.

Comparing to the explicit expressions for $\cH_A$ obtained in Section \ref{casdiffzt}, 
we see that these fermion momenta are functions of differential operators in 
both the eigenvalues $z_i$ and the off-diagonal element $t_0$.
In hermitian matrix models and unitary matrix models, 
the emergent fermions are the eigenvalues of the relevant
matrix. Here, however, the $k=0$ emergent fermions have no such 
direct connection to eigenvalues of $Z$. \newline

\section{Free particle structures and counting on $\cM_N$}\label{genNFacts} 

In this section we extend aspects of our $N=2$ discussion of the algebra of
gauge invariant functions and the rings of scale invariant and gauge invariant
differential operators to the case of general $N$. 

Following our considerations for the $k=0$ sector from Section \ref{casfuns}, 
we show that the momenta of the free fermions are determined 
in terms of differential operators on $ \cM_{N}$. 
We show that the correspondence between 
states of the $k=0$ sector and the unitary matrix model 
(and hence free fermion wavefunctions) extends beyond
two-point functions to a class of three-point functions.

At $N=2$ we have shown that the number of generators of the ring of 
gauge invariant functions on $\cM_N$ will be $N^2 +1$. 
At $N=3$, we will show that there is an interesting twist but that 
the above statement remains true in a refined form. 

Finally we study the $m = n = k$ sector. This is the maximum possible value of $k$, 
in contrast to our studies of $k=0$ which is the minimum possible value.
This sector consists of traces and multi-traces of $\Zd Z$ and we show that 
it may be mapped to $N$ free fermions in a one-dimensional harmonic oscillator potential.
This is a second, distinct appearance of free particles
in complex matrix models.

\subsection{The $k=0$ sector revisited} \label{sec:k=0generalN}

We first observe that our construction of free fermion momenta 
as functions of differential operators in $z_i, t_{ij}$
may be extended to general $N$ in a slightly weaker form as follows.
 
The construction in the previous section may be carried out 
for general $N$ by identifying differential operators which measure 
higher order Casimirs. These will be traces of higher powers of the $G_i$.
We have not found closed form expressions analogous to \eq{eq:c2pi}
for higher $N$ since this would require us to solve arbitrary order polynomials.
However, since the $p_i$ are integer or half-integer, they may always 
be determined in terms of the eigenvalues of the Hamiltonians \cite{Zelobenko:1973},
and hence implicitly in terms of differential operators in $z_i, t_{ij}$.
We have thus identified an implicit map from 
$k=0$ operators to fermions on a circle for all finite $N$. \newline

We next conjecture that the $k=0$ sector may be descibed as the kernel of the 
differential operator $\tr G_2 G_3$.
Let us recall from Section \ref{finitesetgens} that the differential operator
\be
\tr (G_2 + G_3)^2 = \tr ( G_2^2 + 2 G_2 G_3 + G_3^2 )
\ee
measures $C_2(\g)$, and so $\tr G_2 G_3$ measures 
\be \label{eq:G2G3cas}
\frac{1}{2} \left( C_2(\g) - C_2(\a) - C_2(\b) \right).
\ee
Since for a $k=0$ operator $\g = (0, \a, \b)$, we have that
\be
C_2 (\g) = C_2 (\a) + C_2(\b)
\ee
and so 
\be \label{eq:trG2G3is0}
\left( \tr G_2 G_3  \right) \,O^{k=0}(Z, \Zd) = 0.
\ee
As a brief aside, note that the action of the Brauer contraction element $C_{1 \bar 1}$ on $\T{Z}{i}{j} \T{\Zd}{k}{l} $ is as follows \cite{Kimura:2007wy}:
\be
C_{1 \bar 1} \, \left( \T{Z}{i}{j} \T{Z}{k}{l} \right) =  \T{\delta}{i}{l} \T{(\Zd Z)}{k}{j}.
\ee
Since 
\be
\T{(G_2)}{p}{q}  \T{Z}{i}{j}  =   \T{\delta}{i}{q} \T{Z}{p}{j}   \qquad \mathrm{and} \qquad 
- \T{(G_3)}{q}{p}  \T{\Zd}{k}{l}  =   \T{\delta}{q}{l} \T{\Zd}{k}{p} 
\ee
we have 
\be
- \tr G_2 G_3  \left(  \T{Z}{i}{j} \T{\Zd}{k}{l} \right) =  \T{\delta}{i}{l} \T{(\Zd Z)}{k}{j}
\ee
and since $\tr G_2 G_3$ acts via the Leibniz rule, the action of $-\tr G_2 G_3$ on
\be
\cO = \T{Z}{i_1}{j_1} \T{Z}{i_2}{j_2} \cdots \T{Z}{i_m}{j_m} \T{\Zd}{p_1}{q_1} \T{\Zd}{p_2}{q_2} \cdots \T{\Zd}{p_n}{q_n} 
\ee
is that of the sum over all individual contractions
\be
C = \sum_{r=1}^m \sum_{s=1}^n { C_{r\bar{s}} }.
\ee 
Similarly the action of the laplacian 
\be
\sq = \tr \left( \dbyd{Z} \dbyd{Z^{\dagger}} \right)
\ee
on $\T{Z}{i}{j} \T{\Zd}{k}{l}$ is given by
\be \label{eq:lapwick}
\sq \left(  \T{Z}{i}{j} \T{\Zd}{k}{l} \right) =  \T{\delta}{i}{l} \T{\delta}{k}{j}.
\ee
which is a Wick contraction using the two point function (\ref{basic2pt}), and as before extends via the Leibniz rule. It was noted in \cite{Kimura:2007wy} that the $k=0$ operators have no self Wick contractions and so we have
\be \label{eq:lapk=0}
\sq \,O^{k=0}(Z, \Zd) = 0 ~,
\ee
a result we shall use later in Section \ref{sec:SHO}. For now we simply note that it is possible to construct simple examples which 
show that the  $k = 0$ operators do not comprise  the full  
kernel of $\Box $.

We expect however that the converse of (\ref{eq:trG2G3is0}) is true for any $N$
\be \label{eq:G2G3conj}
\tr ( G_2 G_3 ) \, \cO = 0  \qquad \Rightarrow \qquad \cO = \cO^{k=0}
\ee
meaning that the kernel of $\tr ( G_2 G_3 )$ is exactly the $k=0$ sector.
As a differential operator, $ \tr ( G_2 G_3) $ can be viewed as a modification 
of the laplacian which is invariant under scalings of $Z$ and $Z^{\dagger}$.

It is instructive to try and construct  a counterexample
to (\ref{eq:G2G3conj}). 
From (\ref{eq:G2G3cas}) we know that $\tr ( G_2 G_3 ) \, \cO = 0$
is equivalent to 
\be\label{eq:C2s}
C_2(\g) = C_2(\a) + C_2(\b).
\ee
The operator with labels
\be
\a = [1,1], \qquad \b = [1,1], \qquad \g = ( k=1, \g_+ = [1], \g_- = [1] )
\ee
has Casimirs
\be
C_2(\a) = 2, \qquad C_2(\b) = 2, \qquad C_2(\g) = 4
\ee
however this operator in fact does not exist since it fails our 
$N=2$ constraint (\ref{eq:N=2newconstraint}) in the form:
\be
c_1(\a) + c_1(\b) \le N + k.
\ee
This example supports (\ref{eq:G2G3conj}) and shows that it is
sensitive to finite $N$ constraints of the Brauer basis.

\subsection{Three-point functions of $k=0$ operators} \label{sec:threepoint}

In Section \ref{sec:N=2,k=0} we reviewed the map between $k=0$ operators and Unitary matrix model operators, and the result that the two point functions on both sides of the correspondence agree. Here we show that the same is true for a class of three-point functions. The correlators we consider are a subclass of the correlators
\begin{eqnarray}
\langle 
\cO_{A_1}(Z,\Zd) \, \cO_{A_2}(Z,\Zd) \, \cO^{\dag}_{A_3}(Z,\Zd)
\rangle
\end{eqnarray}
where $A_1 =  R_1 \bar{S_1} $ is a short notation for the 
labels of the operators  in the $k=0$ sector, and similarly for $A_2$ and $A_3$. 
 In terms of Brauer projectors the operators are defined by
\be
\cO_{R \bar S}(Z,\Zd) = \tr_{m,n}( P_{R \bar S} Z\otimes \Zd )
\ee
where $P_{R \bar S}$ is defined in Appendix \ref{app:Brauer}. The subclass we study is that in which $m_{1}+m_{2}=m_{3}$ and $n_{1}+n_{2}=n_{3}$. Performing the Wick contractions, we get 
\begin{eqnarray}
\langle 
\cO_{A_1}(Z,\Zd) \, \cO_{A_2}(Z,\Zd) \, \cO^{\dag}_{A_3}(Z,\Zd)
\rangle
=m_3!n_3!\tr_{m_3,n_3}((P_{A_{1}}\circ P_{A_{2}})P_{A}).
\label{threepointCal}
\end{eqnarray}
The calculation is very similar to those in \cite{Corley:2001zk,Corley:2002mj}. It is convenient to express projectors as an integral over   
the $U(N)$ group as 
\begin{eqnarray}
P_{\gamma} &=& Dim\gamma\int dU \chi_{\gamma}(U^{\dagger})U
\end{eqnarray}
where $Dim\gamma$ is the dimension of the $U(N)$ representation $\g$; this follows from Schur-Weyl duality. We can therefore calculate (\ref{threepointCal}) via 
\begin{eqnarray}
\tr_{m,n}((P_{A_{1}}\circ P_{A_{2}})P_{A_{3}}) &=& DimA_{3}\int dU_{3}\chi_{A_{3}}(U_{3}^{\dagger})\tr_{m_3,n_3}((P_{A_{1}}\circ P_{A_{2}})U_{3})\cr
&=&  Dim A_{3}d_{A_{1}}d_{A_{2}} \int dU_{3} \chi_{A_{1}}(U_{3}) \chi_{A_{2}}(U_{3}) \chi_{A_{3}}(U_{3}^{\dagger})  \cr
&=&  Dim A_{3}d_{A_{1}}d_{A_{2}} g(A_{1},A_{2};A_{3}) \cr
&=&  Dim A_{3}d_{R_{1}}d_{R_{2}}d_{S_{1}}d_{S_{2}} g(R_{1},R_{2};R_{3}) g(S_{1},S_{2};S_{3})
\end{eqnarray}
where $d_A \equiv d_{R \bar S } = d_R d_S  $ 
 and the following has been used to get the second equality:
\begin{eqnarray}
\tr_{m,n}((P_{A_{1}}\circ P_{A_{2}})U_{3})
&=&
d_{A_{1}}d_{A_{2}}\chi_{A_{1}}(U_{3})\chi_{A_{2}}(U_{3})
\end{eqnarray}
and 
\begin{eqnarray}
g(A_{1},A_{2};A_{3})=
\int dU_{3}
\chi_{A_{1}}(U_{3})
\chi_{A_{2}}(U_{3})
\chi_{A_{3}}(U_{3}^{\dagger}) 
\end{eqnarray}
is the Littlewood-Richardson coefficient which counts the number of $A_{3}$ in the tensor product $A_{1}\otimes A_{2}$. 
This derivation shows that integration over $ \cM_N$ can be done using Brauer algebras.

\subsection{Finite $N$ counting of single traces and multi-traces}

Since $Z$ and $\Zd$ 
do not in general commute, enumerating multi-trace operators
in one complex matrix is an equivalent problem to 
enumerating multi-trace operators in two hermitian matrix models. 

Therefore an important check on the Brauer basis 
is that the counting of Brauer basis operators agrees
with that of the counting of operators in two-matrix models
\cite{Dolan:2007rq,Brown:2007xh,BCD,Brown:2008ij,Collins:2008gc}.

At finite $N$, we denote the counting of multi-trace operators built from two matrices
with $m$ of one type and $n$ of the other as $Q_{mt}^N ( m , n)$. 
The following two expressions determine $Q_{mt}^N ( m , n)$ in terms of 
group theoretical quantities
\cite{Willenbring:2005,Dolan:2007rq,Brown:2007xh}:
\bea\label{Qmtformulae} 
Q_{mt}^N ( m , n) 
 & = &  \sum_{ \substack { R \vdash m+n , \Lambda \vdash m+n \\ 
                            c_1 ( R ) \le N , c_1 ( \Lambda ) \le 2 } }
 C ( R , R , \Lambda )    g ( [m] ,[n] ;  \Lambda ) 
\eea 
and
\bea\label{Qmtformulae2} 
Q_{mt}^N ( m , n)  & =  & \sum_{ \substack { R \vdash m + n \\  c_1 ( R ) \le N }  } 
\sum_{ \substack{ R_1 \vdash m \\  c_1 ( R_1) \le N  } }
 \sum_{ \substack{ R_2  \vdash n \\  c_1 ( R_2) \le N   } } 
     g ( R_1 , R_2 ;  R )^2       .
\eea 
Here $ C ( R , R ; \Lambda ) $ is the multiplicity 
of the irreducible representation $ \Lambda $ of $ S_{m+n} $ appearing 
in the tensor product of irreducible representations $ R \otimes R  $ 
of $ S_{m+n}$, and $g(\cdot \, ,\cdot\, ; \cdot)$ is a Littlewood-Richardson coefficient.

Defining the finite $N$ multi-trace generating function
\bea
Z_{mt}^N ( x , y ) &=& \sum_{ m , n }{ Q_{mt}^N ( m , n)  ~ x^m y^n } ,
\eea
the relation between the counting of single-traces and multi-traces is given by the Plethystic Logarithm \cite{Benvenuti:2006qr,Feng:2007ur}:
\bea \label{eq:plog}
Z_{st}^N ( x , y ) &=& \sum_{k=1}^\infty  \frac{\mu(k)}{k} \log \left( Z_{mt}^N ( x^k , y^k ) \right) .
\eea
At $N=2$ this leads to the five single traces identified in equation
 (\ref{eq:N=2ring}).

 At $N=3$ the Plethystic Logarithm gives the single-trace generating function
\bea
\sum_{ m , n } Q_{st}^{N=3}( m ,n ) \, x^m y^n  
 &=&  1 + x + y + x^2 + y ^2 + xy + x^3 +  y^3 \nn \\			
 & &  + \, x^2 y + x y^2 + x^2 y^2 + x^3 y^3 - x^6 y^6
\eea
The interpretation of this generating function is that there are 11 independent single trace operators along with one \textit{syzygy} (algebraic relation) which occurs at order $(m,n) = (6,6)$. It would be interesting to find this relation explicitly.

We expect that the reduction multiplicities for 
$ B_{N}( m ,n ) $ to $ S_m \times S_n $, which we have denoted by $ M^{ \gamma ; N }_{ \alpha , \beta }  $, satisfy the counting
\bea \label{expectcount}  
\sum_{ \gamma , \alpha, \beta } \left( M^{ \gamma  ; N }_{ \alpha , \beta }   \right)^2 &=& Q_{mt}^N ( m , n).
\eea 
We show in the next section
that this is true for $N>m+n$ and it is also
consistent with our calculations at $N=2$ in Section \ref{countN=2}. 
It would be interesting to prove it for general $N$, with or without 
using the connections to matrix models.

\subsection{Large $N$ Brauer basis counting}

For  $N$ sufficiently large, i.e $N > m +n $, 
we denote the counting of multi-trace operators by $Q_{mt}(m,n)$. 
 The formulae in (\ref{Qmtformulae}), (\ref{Qmtformulae2})
 give rise to this quantity $Q_{mt}(m,n)$ when $ N > m+n$. 
By P\'{o}lya counting $Q_{mt}(m,n)$ is also given by (see \cite{Dolan:2007rq,Brown:2007xh} \& refs within)
\be
\prod_{r=1}^{\infty}{ \frac{1}{ 1-(x^r + y^r) }} = \sum_{m, n = 0}^{\infty} Q_{mt}(m,n) x^m y^n. 
\ee
In equation (177) of  \cite{Brown:2007xh} the following expression was derived:
\be \label{eq:Nmn-cl}
Q_{mt}(m,n) = \sum_{ c_l (1): \sum_{ l} c_l (1) l = m } ~  \sum_{ c_l (2): \sum_{ l} c_l (2) l = n }
~
\prod_{l}\frac{ (c_{l}(1)+c_{l}(2 )) ! } {c_{l}(1)!c_{l}(2)!}  
\ee
The counting of Brauer basis operators is denoted $N_{sb}(m,n)$ and was shown in \cite{Kimura:2007wy} to be given by 
\begin{eqnarray} \label{eq:Nsb}
N_{sb}(m,n) &=& \sum_{\gamma,A} \left( M_{A}^{\gamma} \right)^{2}. 
\end{eqnarray}
In \cite{Kimura:2007wy} it was argued that this formula correctly 
counts  multi-traces at large $N$, namely that
\bea
N_{sb}(m,n) &=& Q_{mt}(m,n).
\eea
In Appendices \ref{brauerbasiscounting} and \ref{brauerbasiscounting2} we give two proofs of this fact,
firstly by direct comparison to (\ref{eq:Nmn-cl}) and secondly by enumerating invariants in the reduction $GL(N) \times GL(N) \rightarrow  GL(N)$.

\subsection{The $m = n = k$ sector: Operators and free fermions   }  

We recall from the discussion at the start of Section \ref{casfuns} that the integer $k$ is directly related to the minimum number of Brauer contractions involved in the terms which are summed to make up an operator in the Brauer basis.

For $m=n=k$, all terms in an operator involve the maximum number of contractions, which translates into
the fact that these operators are multi-traces of the matrix $Y = \Zd Z$. Since $Y$ is hermitian we find 
the $N$ fermions of the hermitian matrix model emerging in this sector, as follows.

In this sector we have $\g = (k=m, \g_{+} = \emptyset, \g_{-} = \emptyset)$ and $\a = \b$, so the projectors $Q^{\g}_{\a,\b}$ (defined in Appendix \ref{app:Brauer}) are in this sector labelled by $\a$ alone. We write
\be
P_{\a}^{k=m} = Q^{\g}_{\a,\a}  \qquad \mathrm{with} \,\, \g \,\, \mathrm{as} \,\, \mathrm{above}.
\ee
The projector is written in terms of the $k$-contraction operator $C_{(k)}$ defined by 
\bea
C_{(k)}
&=&\sum_{\sigma\in S_{k}}
C_{\sigma(1)\bar{1}}\cdots 
C_{\sigma(k)\bar{k}},
\eea
and the projector $p_{\alpha}$ which projects the holomorphic half of $V^{\otimes k}\otimes \bar{V}^{\otimes k}$ to the representation $\a$. It is proved in Appendix \ref{app:k=mproofs} that the projector takes the form
\begin{eqnarray}
P^{k=m}_{\alpha}=\frac{d_{\alpha}}{k!Dim\alpha}C_{(k)}p_{\alpha}
\label{k=mprojectorinmm}
\end{eqnarray}
and that the operator satisfies the following required properties:
\be \label{projectork=m2}
(P^{k=m}_{\alpha})^{2}=P^{k=m}_{\alpha} \qquad \mathrm{and} \qquad \tr_{k,k}(P^{k=m}_{\alpha})=(d_{\alpha})^{2}
\ee
where $d_{\a}$ is the dimension of the $S_k$ representation $\a$. The operators in the $m=n=k$ sector therefore take the explicit form: 
\begin{eqnarray}
&&
\tr_{k,k}(P_{\alpha}^{k=m}Z^{\otimes k}\otimes Z^{\ast \otimes k})
\cr
&=&
\frac{d_{\alpha}}{k!Dim\alpha}
\tr_{k,k}(C_{(k)}p_{\alpha}Z^{\otimes k}\otimes Z^{\ast \otimes k})
\cr
&=&
\frac{d_{\alpha}}{k!Dim\alpha}
\sum_{\sigma\in S_{k}}
\tr_{k,k}(\sigma 
C_{1\bar{1}}\cdots C_{k\bar{k}}
\sigma^{-1}
p_{\alpha}Z^{\otimes k}\otimes Z^{\ast \otimes k})
\cr
&=&
\frac{d_{\alpha}}{Dim\alpha}
\tr_{k,k}(
C_{1\bar{1}}\cdots C_{k\bar{k}}
p_{\alpha}Z^{\otimes k}\otimes Z^{\ast \otimes k})
\cr
&=&
\frac{d_{\alpha}}{Dim\alpha}
\tr_{k}(
p_{\alpha}Y^{\otimes k})
\end{eqnarray}
where $Y=Z^{\dagger} Z$.
So operators in the $m=n=k$ sector are Schur polymonials constructed from 
$Y$. 

We may understand these results in the following way. First observe that 
$H_L$ annihilates $\T{(Z^{\dagger} Z)}{i}{j}$, since $ H_L = G_2 + G_3$ 
generates the $U(N) $ action on the lower index of $ Z^{\dagger}$ and the 
upper index of $Z$,
\be \label{eq:GLaction}
Z \to UZ, \qquad \Zd \to \Zd \Ud
\ee
and that the product $ \T{(Z^{\dagger} Z)}{i}{j}$
is invariant under this action.  
Traces of powers of  $Y$ are thus also invariant
under (\ref{eq:GLaction}). 

$H_L$ measures $ C_2 ( \gamma )$ which implies that
$ C_2 ( \gamma )= 0  $ for all operators built from $\T{Y}{i}{j}$.
This is consistent with the fact that in the $m=n=k$ sector
$\g = (k=m, \g_{+} = \emptyset, \g_{-} = \emptyset)$ and so $C_2(\g) = 0$. 
We can consider a Casimir of the form 
$ \tr ( Y { \d \over \d Y}  )^2  $ which measures the 
labels of the Young diagram.

By the map discussed in Section \ref{sec:HMM}, Schur polynomials in a hermitian
matrix correspond to the states of $N$ free fermions in a harmonic oscillator potential.
The harmonic oscillator fermions observed here are a second emergence of free 
particles, distinct from those of the $k=0$ sector.

\section{Applications to integrable quantum mechanics} \label{sec:SHO}

\subsection{Review of matrix harmonic oscillator quantum mechanics}

We now return to the dimensional reduction of $\cN = 4$ Super Yang-Mills on $R \times S^3$ in the zero coupling limit, 
truncated to the sector of one complex matrix.
As noted in Section \ref{sec:GaugedMM} one may choose the $A_0 =0 $ gauge 
while imposing Gauss's Law,
yielding the quantum mechanics for the matrix $Z(t)$ defined by the following action
\cite{Hashimoto:2000zp}:
\begin{eqnarray}
\cS = \int dt \tr \bigg( \dot{Z}\dot{Z}^{\dagger}-ZZ^{\dagger} \bigg)
\end{eqnarray}
It is well known that the holomorphic sector of the theory is equivalent to a system of non-interacting fermions in a one-dimensional harmonic oscillator potential \cite{Corley:2001zk,Berenstein:2004kk,Takayama:2005yq}. 
As a subsector of $\cN = 4$ Super Yang-Mills extremal correlators in this sector are protected by supersymmetry \cite{Eden:2000qp,Eden:2000gg} and the states of this sector are dual to the LLM supergravity geometries \cite{Lin:2004nb}.

Going beyond the holomorphic sector, 
we no longer have non-renomalization theorems 
so the connection to supergravity is not straightforward. 
Based on the following investigations,
we will infer properties of any candidate
 string dual  of the  complex matrix model sector 
at zero coupling in Section \ref{sec:Summary}.

We first review the previous analysis of the above theory \cite{Corley:2001zk}. 
The momenta conjugate to $\T{Z}{i}{j}$ and $\T{\Zd}{i}{j}$  
are  
\begin{align}
\T{\Pi}{j}{i} \equiv \Pi_{ \T{Z}{i}{j}} = {\partial L \over \partial  \T{\dot Z}{i}{j} } &=   \T{ \dot{Z}^{\dag}}{j}{i} , &  
\T{\Pi^{\dag}}{j}{i} \equiv \Pi_{ \T{\Zd}{i}{j}} = {\partial L \over \partial  \T{\dot{Z}^{\dag}}{i}{j} }   &=  \T{\dot Z}{j}{i}  . 
\end{align}
The equal time canonical commutation relations are 
\begin{align}  \label{cancom} 
\left[ \T{Z}{p}{q}  , \T{\Pi}{j}{i}  \right] &= i \, \T{\delta}{j}{q}  \T{\delta}{p}{i}   &
\left[ \T{\Zd}{p}{q}  , \T{\Pi^{\dag}}{j}{i}  \right] &= i \, \T{\delta}{j}{q}  \T{\delta}{p}{i} 
\end{align}
so we can identify the conjugate momenta with matrix derivatives in the usual way using (\ref{eq:matrixderiv}). 
We define the creation and annihilation operators:
\begin{align}  \label{aadaggerops} 
A^{\dagger} &=  \frac{1}{\sqrt{2}}(Z-i\Pi^{\dagger}) = \frac{1}{\sqrt{2}}\left(Z-\dbyd{\Zd}\right)   &
A = \frac{1}{\sqrt{2}}(\Zd + i \Pi) &= \frac{1}{\sqrt{2}} \left(\Zd+\dbyd{Z}\right) \cr
B^{\dagger} &=  \frac{1}{\sqrt{2}}(\Zd-i\Pi)  = \frac{1}{\sqrt{2}}\left(\Zd-\dbyd{Z} \right)   &
B =  \frac{1}{\sqrt{2}}(Z+i\Pi^{\dagger}) &= \frac{1}{\sqrt{2}}\left(Z+\dbyd{\Zd} \right) 
\end{align}
Importantly, the dagger  on $\Ad$ does \textbf{not} signify hermitian conjugate of $A$. It signifies purely that this is a creation operator. 
The hermitian conjugate of $\T{\Ad}{i}{j}$ is $ \T{A}{j}{i}$.
The canonical commutation relations become
\begin{align}
[ \T{A}{i}{j} , \T{\Ad}{k}{l} ] &= \T{\delta}{i}{l}\T{\delta}{k}{j}  &
[ \T{B}{i}{j} , \T{\Bd}{k}{l} ] &= \T{\delta}{i}{l}\T{\delta}{k}{j}.
\end{align}
The Hamiltonian and $U(1)$ current take the form
\bea
\hat H &=& \tr \left( - \frac{\d^2}{\d Z \d \Zd} + Z \Zd \right) ~=~ \tr(A^{\dagger}A+B^{\dagger}B)+N^{2} \cr
\hat J &=& \,\,  \tr \left(    Z \dbyd{Z} - \Zd \dbyd{\Zd}     \right) \,  ~=~ \tr(A^{\dagger}A-B^{\dagger}B)
\eea
where $N^{2}$ is the zero point energy for $N^{2}$ harmonic oscillators in two dimensions.

The ground state of this system satisfies $A \vac = B \vac =0$. The corresponding (non-normalised) wavefunction $\Psi_{0}=\langle Z,\bar{Z}|0\rangle$ is
\begin{eqnarray}
\Psi_{0}(Z,\Zd) = e^{-\tr(ZZ^{\dagger})}.
\end{eqnarray}
Holomorphic gauge invariant excitations of this system are defined by the constraint $B| \cO \rangle=0$ and consist of operators built from $A^{\dagger}$ acting on the ground state. These may be written as
\begin{eqnarray}
 \tr_{n}(\sigma (A^{\dagger})^{\otimes m}) \vac
\label{tracebasisAdagger}
\end{eqnarray}
where $\sigma$ is an element of $S_{n}$, and controls how the indices are contracted to form either a single or multi-trace operator. A more convenient basis for operators of the form (\ref{tracebasisAdagger}) is the Schur polynomial basis (for details see \cite{Corley:2001zk}):
\begin{eqnarray}
\ket{\Psi_R} &=& \chi_{R}(\Ad) \vac
\label{schurpolynomialholobasis}
\end{eqnarray}
where $\chi_{R}$ is the character of the $U(N)$ representation $R$. Since 
\bea
\Ad e^{-\tr(ZZ^{\dagger})} &=& \sqrt{2}Z e^{-\tr(ZZ^{\dagger})},
\eea
we may write 
\begin{eqnarray}
\Psi_R (Z,\Zd) &=& \chi_{R}(\sqrt{2} Z)e^{-\tr(ZZ^{\dagger})}.
\label{root2Z}
\end{eqnarray}
This state has $E=m+N^{2}$ and $J=m$ and is holomorphic in $Z$ up to
the exponential factor. If we triangularize $Z$ and redefine the wavefunction by absorbing the Jacobian of the transformation into the definition of the wavefunction, it becomes a wavefuction for $N$ fermions 
in the Lowest Landau Level of the Quantum Hall system \cite{Corley:2001zk,Berenstein:2004kk,Ghodsi:2005ks}. \newline

\subsection{Non-holomorphic sector} \label{sec:nonholo}

The most general eigenstate can be constructed by acting with both $A^{\dagger}$ and $B^{\dagger}$ on the ground state,
\bea \label{eq:Psidef}
\ket{\Psi_{ \cO } } &=& \cO(A^{\dagger}, B^{\dagger}) \, \vac
\eea
where $\cO(A^{\dagger}, B^{\dagger})$ is a gauge invariant polynomial
constructed from $m$ $A^{\dagger}$'s and $n$ $B^{\dagger}$'s. 

 The wavefunction of such a state may be written as
\begin{eqnarray}
\Psi_{\cO} (Z,\Zd) &=&  \braket{Z,\Zd}{\Psi_{\cO} }  ~=~  \cO(A^{\dagger}, B^{\dagger}) \, e^{-\tr(ZZ^{\dagger})}.
\end{eqnarray}
The Brauer Algebra may be used to organise the states above. Such states are analogous to those used in Section \ref{casfuns} and take the form
\bea
\ket{\Psi^{\g}_{\a,\b;i,j}} &=& \cO^{\g}_{\a,\b;i,j} (\Ad,\Bd) \vac
\eea
where the labels are explained in Appendix \ref{app:Brauer}. This state has $E=m+n+N^{2}$ and $J=m-n$.

Unlike for the holomorphic sector wavefunctions, we have
\be
\cO(A^{\dagger}, B^{\dagger})e^{-\tr(ZZ^{\dagger})} \neq \cO(\sqrt{2}Z, \, \sqrt{2}\Zd) \, e^{-\tr(ZZ^{\dagger})}
\ee
because the derivative of $Z$ inside $A^{\dagger}$ acts on $Z$ which comes from the action of $B^{\dagger}$ on the exponential factor. For example we have
\be
\tr(A^{\dagger}B^{\dagger})e^{-\tr(ZZ^{\dagger})} =  \left( 2\tr Z Z^{\dagger} - N^2 \right) e^{-\tr(ZZ^{\dagger})}
\ee
and in general the correct relation is 
\bea 
\fbox{$ \quad 
\Psi_{\cO} (Z,\Zd) = \cO(A^{\dagger}, B^{\dagger})e^{-\tr(ZZ^{\dagger})} 
                 = \left[ e^{-\frac{\sq}{2}} \cO(\sqrt{2}Z, \, \sqrt{2}\Zd) \right] e^{-\tr(ZZ^{\dagger})} \quad \label{eq:exp-boxover2}
$} 
\eea
where $\sq$ is the laplacian $\tr \dbyd{Z} \dbyd{\Zd}$ and the brackets indicate that the derivatives in $\sq$ act only on $\cO(\sqrt{2}Z, \, \sqrt{2}\Zd)$ and not on the exponential. $e^{-\frac{\sq}{2}}$ is defined by its series expansion; it was observed in (\ref{eq:lapwick}) that the laplacian generates Wick contractions and so here $e^{-\frac{\sq}{2}}$ performs
a normal ordering, subtracting terms in which pairs of $ \sqrt{2}Z $ and $  \sqrt{2}Z^{\dagger}$ 
have been contracted (c.f. \cite{Polchinski:1998ordering}).

Note however that in a $k=0$ operator we have from (\ref{eq:lapk=0}) that
\be
\sq \, \cO^{k=0} = 0
\ee
and so we can replace $A^{\dagger}$ and $B^{\dagger}$ with $\sqrt{2} Z$ and $\sqrt{2} \Zd$ respectively without worrying about the above subtlety.

We can define operators corresponding to the $G_i$  in (\ref{eq:G_a}) as follows.
\begin{align}
\T{(\hat{G}_1)}{i}{j} &~=~ (B^{\dag}B)^{i}_{\,j}   &   \T{(\hat{G}_2)}{i}{j} &~=~ (A^{\dag}A)^{i}_{\,j}  \nn  \\
\T{(\hat{G}_3)}{i}{j} &~=~ -B^{\dag}{}^{k}_{\,\,j} B^{i}_{\,k}   & \T{(\hat{G}_4)}{i}{j} &~=~ -A^{\dag}{}^{k}_{\,\,j} A^{i}_{\,k}
\end{align}
Defining $\ket{\T{A}{i}{j}} = \T{A}{i}{j} \vac$ and so on, using the commutation relations we find
\begin{align}
\T{(\hat{G}_1)}{i}{j} \ket{ \T{B^{\dag}}{p}{q} } &~=~  \de{p}{j} \ket{ \T{B^{\dag}}{i}{q} } &
\T{(\hat{G}_2)}{i}{j} \ket{ \T{A^{\dag}}{p}{q} } &~=~  \de{p}{j} \ket{ \T{A^{\dag}}{i}{q} } \nn \\
\T{(\hat{G}_3)}{i}{j} \ket{ \T{B^{\dag}}{p}{q} } &~=~ -\de{i}{q} \ket{ \T{B^{\dag}}{p}{j} } &
\T{(\hat{G}_4)}{i}{j} \ket{ \T{A^{\dag}}{p}{q} } &~=~ -\de{i}{q} \ket{ \T{A^{\dag}}{p}{j} } \label{eq:Ghataction}
\end{align}
which is the same as the adjoint action of the operators $G_i$ defined in (\ref{eq:G_a}) on the matrices $Z, \Zd$ (see equation (11) of \cite{Kimura:2008ac}).

The result is that we can define harmonic oscillator Casimir operators
\be 
 \hat \cH_A =  \Big \{ \hat H_1,\quad  \hat H_2,\quad  \hat{ \bar {H}}_1 , \quad \hat{ \bar {H}}_2,\quad  \hat  H_L \Big \} 
\ee
by replacing  $ G_i $ in (\ref{eq:Hamiltonians}) with $\hat{G}_i$. 
The eigenvalues of hatted Casimirs  acting on
 $ \cO^{\g}_{\a , \b ; i, j } ( A^{\dagger} , B^{\dagger } ) |0\rangle  $
 are the same as those of the corresponding unhatted Casimirs 
 acting on   $ \cO^{\g}_{\a , \b ; i, j } (  Z , Z^{\dagger} ) $. 
This is because the same commutator manipulations can be done to 
evaluate both, and the arguments which prove that 
 $ \cO^{\g}_{\a , \b ; i, j } (  Z , Z^{\dagger} ) $ are eigenstates of 
the Casimirs in (\ref{eq:Hamiltonians}) 
 also prove that  $ \cO^{\g}_{\a , \b ; i, j } ( A^{\dagger} , B^{\dagger } ) |0\rangle  $ are eigenstates of the hatted versions. 
 
We can take this one step further. Noting that
\bea
\left[  \T{Z}{i}{j} , - \frac{\sq}{2} \right]  &=& 
 \frac{1}{2} \left( \dbyd{\Zd}\right)^{\!\! i}_{j}  \\
\Rightarrow \left[  \T{Z}{i}{j} , e^{- \frac{\sq}{2}} \right]  &=&  
 \frac{1}{2} \left( \dbyd{\Zd} \right)^{\!\! i}_{j} e^{- \frac{\sq}{2}}
\eea
and similarly
\bea
\left[ \T{\Zd}{i}{j}  , e^{- \frac{\sq}{2}} \right]  &=&  
 \frac{1}{2} \left( \dbyd{Z} \right)^{\!\! i}_{j} e^{- \frac{\sq}{2}}
\eea
then using (\ref{eq:exp-boxover2}) we derive 
\bea
\T{\Ad}{i}{j} \, \Psi_{\cO} (Z,\Zd) &=&  
     \T{\Ad}{i}{j} \, \cO(\Ad,\Bd) \, e^{- \tr(ZZ^{\dagger})}    \cr
&=&   \T{\Ad}{i}{j}  \left[ e^{-\frac{\sq}{2}} \cO(\sqrt{2}Z, \, \sqrt{2}\Zd) \right] e^{-\tr(ZZ^{\dagger})} \cr
&=&  \left[ e^{-\frac{\sq}{2}}  \left( \sqrt{2}\T{Z}{i}{j} \right) \cO(\sqrt{2}Z, \, \sqrt{2}\Zd) \right] e^{-\tr(ZZ^{\dagger})} 
\eea
where again the brackets indicate that the derivatives act only on $\cO(\sqrt{2}Z, \, \sqrt{2}\Zd)$ and not on the exponential. Similarly
\bea
\T{A}{i}{j} \, \Psi_{\cO} (Z,\Zd) &=& 
 \left[ e^{-\frac{\sq}{2}} \left( \frac{1}{\sqrt{2}}\left( \dbyd{Z} \right)^{\!\! i}_{j} \right)  \cO(\sqrt{2}Z, \, \sqrt{2}\Zd) \right] e^{-\tr(ZZ^{\dagger})}   
\eea
implying the following relation between $\hat{G}_2$ and $G_2$:
\bea
\T{(\hat{G}_2)}{i}{j} \, \Psi_{\cO} (Z,\Zd) &=&
 \left[ e^{-\frac{\sq}{2}} \, \T{({G}_2)}{i}{j} \,  \cO(\sqrt{2}Z, \, \sqrt{2}\Zd) \right] e^{-\tr(ZZ^{\dagger})} \label{eq:AdAij}
\eea
Similar results apply to the remaining $\hat{G}_i$, the Hamiltonians $\hat H_i$ as well as the canonical Hamiltonian 
\be
\hat H ~=~ \hat H_1  + \hat{\bar{H}}_{1} + N^2 ~=~ \tr ( A^{\dagger } A + B^{\dagger}  B  ) + N^2
\ee
whose action on wavefunctions $ \Psi(Z,\Zd)$ can be written in terms of the (first-order) scaling operator $H$:
\be \label{eq:H}
H ~=~ H_1 + \bar H_1 +N^2 ~=~  \tr\left(Z { \d \over \d Z }  + Z^{\dagger} { \d \over \d Z^{\dagger} }\right) +N^2 .
\ee 
Applying (\ref{eq:AdAij}) and the corresponding relation for $\hat{G}_3$ we find that
\bea
\fbox{$\quad 
\hat H  \Psi_{\cO} (Z,\Zd)  =  \left[ e^{-\frac{\sq}{2}}  \, H \, \cO(\sqrt{2}Z, \, \sqrt{2}\Zd) \right] e^{-\tr(ZZ^{\dagger})}. \label{eq:ABZZd}
\quad $}
\eea
A similar manipulation in the holomorphic sector was performed in Appendix A of \cite{Yoneya:2005si}. Note that for a $k=0$ operator we have $\sq \, \cO^{k=0} = 0$ and so the above analysis gives
\bea\label{hatHH}  
\hat H\left[  \cO^{k=0}(\Ad,\Bd) \, e^{- \tr(ZZ^{\dagger})} \right] =  \left[  H \, \cO^{k=0}(\sqrt{2}Z, \, \sqrt{2}\Zd) \right] e^{-\tr(ZZ^{\dagger})}. 
\eea

The inner product on wavefunctions may be derived using
\be \label{eq:completenessZ}
\int [dZ d\Zd] \, \ket{Z,\Zd}\bra{Z,\Zd} = 1
\ee
where $[dZ d\Zd] = \prod_{i,j}{ d Z_{ij} d \Zd_{ij}}$, as follows:
\bea
\braket{\Psi_{\cO_1} }{\Psi_{ \cO_2} }
&=& \frac{1}{\pi^{N^2}} \int {[dZ d\Zd] \, \braket{\cO_1(\Ad, \Bd)}{Z,\Zd} \braket{Z,\Zd}{\cO_2(\Ad, \Bd)} } \cr
&=& \frac{1}{\pi^{N^2}} \int {[dZ d\Zd] \,
\overline{\Psi_{ \cO_1} (Z,\Zd)} \Psi_{ \cO_2} (Z,\Zd) }  \label{eq:AB-IP}
\eea
where $\pi^{N^2}$ compensates for using non-normalised wavefunctions, and is found by imposing
\be
\braket{\Psi_0}{\Psi_0} = 1.
\ee
Using (\ref{eq:exp-boxover2}), the above expression (\ref{eq:AB-IP}) becomes
\bea
\braket{\Psi_{ \cO_1} }{\Psi_{ \cO_2} } &=& \frac{1}{\pi^{N^2}} \int {[dZ d\Zd] \, 
\overline{\cO_1(\Ad, \Bd) e^{-\tr Z \Zd}} \cO_2(\Ad, \Bd) e^{-\tr Z \Zd} } \cr
&=& \frac{1}{\pi^{N^2}} \int { [dZ d\Zd] \, 
\overline{ \left( e^{-\frac{\sq}{2}} \cO_1(\sqrt{2} Z,\sqrt{2} \Zd) \right) } 
\left( e^{-\frac{\sq}{2}} \cO_2(\sqrt{2} Z,\sqrt{2} \Zd) \right) e^{-2\tr Z \Zd} } \cr && \label{eq:AB-IP1}
\eea
and rescaling factors of two we have the result
\bea 
\braket{\Psi_{ \cO_1} }{\Psi_{ \cO_2} } = \frac{1}{(2\pi)^{N^2}} \int { [dZ d\Zd] \, 
\overline{ \Big( e^{-\sq} \, \cO_1(Z,\Zd) \Big) } 
\Big( e^{-\sq} \, \cO_2(Z,\Zd) \Big) e^{-\tr Z \Zd} }  \label{eq:AB-IP2} 
\eea
which is the non-holomorphic generalisation of (A.12) of \cite{Yoneya:2005si}.  

In the next section we use the right hand side of the above equation to define an inner product on gauge invariant polynomials $\cO(Z,\Zd)$
rather than the harmonic oscillator wavefunctions $\Psi_{\cO} (Z,\Zd)$ which contain exponentials.

\subsection{Related integrable quantum mechanics models}  \label{sec:SHO3}

In the discussion above we related the action of the Hamiltonian $\hat H$ in terms of the (first-order) scaling operator $H$:
We thus have an explicit map (\ref{eq:ABZZd}) relating the the action of $\hat H$ on its eigenstates 
\bea  \label{eq:states1}
\Psi_{\cO} (Z,\Zd)  &=&  \cO(\Ad,\Bd) \, e^{- \tr(ZZ^{\dagger})}
\eea
to the action of $H$ on its eigenstates
\be  \label{eq:states2}
\cO(Z,\Zd).
\ee

The right hand side of (\ref{eq:AB-IP2}) can be used to define an inner product on polynomial functions of 
$ Z , Z^{\dagger} $. Explicitly this inner product is 
\bea
( \cO_1(Z,\Zd) , \cO_2(Z,\Zd) ) &=& \frac{1}{(2\pi)^{N^2}} \int { [dZ d\Zd] \, 
\overline{ \Big( e^{-\sq} \, \cO_1(Z,\Zd) \Big) } 
\Big( e^{-\sq} \, \cO_2(Z,\Zd) \Big) e^{-\tr Z \Zd} } . \cr && \label{eq:ZZd-IP}
\eea
Note that the two inner products $(\,\cdot\, , \,\cdot\,)$ and $\braket{\,\cdot\,}{\,\cdot\,}$ 
are defined on two different Hilbert spaces:
\bea
(\,\cdot\, , \,\cdot\,) &:& \left\{ \mathrm{Polynomials~in~} Z, \Zd \right\} \to \R   \cr
\braket{\,\cdot\,}{\,\cdot\,} &:& \big\{ \mathrm{Harmonic~oscillator~states} \big\} \to \R.
\eea
By construction the inner products satisfy 
\bea
( \cO_1(Z,\Zd) , \cO_2(Z,\Zd) )  &=& \braket{\Psi_{ \cO_1} }{\Psi_{ \cO_2} }
\eea
and each inner product is diagonalised by the corresponding Brauer basis 
\be
\cO^{\g}_{\a,\b;i,j}(Z,\Zd)  \qquad   \mathrm{and}  \qquad \ket{\Psi^{\g}_{\a,\b;i,j}} = \cO^{\g}_{\a,\b;i,j}(\Ad,\Bd) \vac .
\ee
The relevant Hamiltonians have the properties that 
\begin{itemize}
\item $\hat{\cH}_A$ are hermitian with the inner product $\braket{\,\cdot\,}{\,\cdot\,}$
\item $\cH_A$ are hermitian with the inner product $(\,\cdot\, , \,\cdot\,)$
\end{itemize}
The inner product $(\,\cdot\, , \,\cdot\,)$ can in fact be constructed 
by starting from the inner product arising from the zero-dimensional 
complex matrix model and requiring $H_i$ to be hermitian.
This involves subtracting $ Z , Z^{\dagger} $ 
contractions and is discussed in Appendix \ref{sec:innprod}.

In addition to the original Hamiltonian $\hat H$,  
it is natural to consider the  conserved charges 
e.g $ \tr \hat G_2^2 $ as Hamiltonians. 
Since these {\it higher} Hamiltonians were constructed 
to be simultaneously diagonalised with $\hat H$ in   the
Brauer algebra basis, we know these are solvable 
Hamiltonians related to the Brauer algebra. 
Related to these higher Hamiltonians are simpler 
ones which are obtained by replacing that hatted $G$'s 
with unhatted ones. For example $ H_2 = \tr G_2^2 $ or 
$ H_L = \tr ( G_2 +G_3)^2 $ 
define  solvable quantum mechanics models. 
They are second order in derivatives as opposed to fourth 
 order like $ \hat H_2 , \hat H_L $, and hermitian in the 
 inner product (\ref{eq:ZZd-IP}).
 
 Considering the expressions in (\ref{eq:N=2ztCasimirs}) 
 and comparing with equation (2.4) of \cite{Pasquier:1994cs}
 we see that  these integrable quantum mechanics 
 models are non-holomorphic 
 generalizations of  the Calogero-Sutherland model 
 at a specific coupling 
 (see also \cite{Agarwal:2006nv,Polychronakos:2006nz} 
 for related literature). 
 A natural question is whether the Calogero-Sutherland model at generic 
 coupling has such integrable non-holomorphic generalizations.  
 Although we have not written out the Hamiltonians  
 for general $N$ explicitly as differential operators 
 in terms of coordinates on $ \cM_N$ it is clear that this 
 can be done by changing variables from $ gl( N ,  \mC ) $
to $ z_i, t_{ij}$.

\section{Summary and outlook}  \label{sec:Summary}

We described free particle structures hidden in matrix models of an $N \times N$
complex matrix $Z$. We related these structures to the geometry of the 
 configuration space $\cM_N$ of gauge-inequivalent
configurations, a space of  dimension $N^2 + 1$.
We showed that $\cM_N$ supports an interesting  class of functions, obtained 
from the gauge invariant functions of $Z$.
The Schur decomposition gives coordinates $z_i , t_{ij}$ 
useful for describing $ \cM_N$. Integrals over 
complex matrices give a measure of integration on $\cM_N$
which can be used to define an inner product 
on gauge invariant functions of $z_i , t_{ij}$.   
Following \cite{Kimura:2007wy}, Brauer algebras  $B_N( m , n) $ 
give orthogonal bases which diagonalise the inner products which arise. 
Higher Casimirs constructed from the Brauer 
algebra, which  resolve these orthogonal bases \cite{Kimura:2008ac}, 
give rise to a complete set of scale and gauge 
invariant differential operators on $\cM_N$. 

Among the labels of the Brauer basis is a non-negative integer $k$. 
For any $N$ the $k=0$ sector has states in one-to-one correspondence with 
those of $N$ free fermions on a circle. These states
correspond to the composite representations $ R \bar S $ 
which play a role in two dimensional Yang Mills.  
The differential operators which measure Casimirs
are polynomials in the free fermion momenta; 
for $N=2$ we inverted these relations to write the 
momenta as algebraic functions of the differential operators. 
We conjectured that the $k=0$ sector is the kernel of a scale invariant version of the laplacian, 
the operator  $\tr ( G_2 G_3 )$. We also gave an  equality of correlators 
between the unitary matrix model and the complex matrix model 
for a class of three-point functions of $k=0$ operators. 
It is important to note that while the usual emergence of 
free fermions in matrix models can be seen from a change of variables 
to eigenvalues \cite{Mehta}, here the $k=0$ sector has no direct relation to 
the eigenvalues of $Z$. Indeed the operator $ \tr ( G_2 G_3 )$ 
characterizing it involves derivatives with respect to $ z_i $ as well 
as the off-diagonal $t_{ij}$. 

Another interesting sector where states are counted by  Young diagrams is 
the sector $ m = n = k $. This is a sector of gauge invariant 
functions of $ \T{(Z^{\dagger}Z )}{i}{j}$ which is the kernel of 
another second order operator on $ \cM_N $, namely $\tr G_L^2$ 
as defined above (\ref{eq:Hamiltonians}).
We observe that $k$ appears to interpolate between 
radial and angular free particle systems on a plane.
It would be interesting to further elucidate 
this in a stringy context.

A precise understanding of the commutative 
ring of scale and gauge invariant differential 
operators led us to computational results on the 
reduction multiplicities of representations of 
$ B_N ( m , n )  $ to $ S_m \times S_n$ for $N=2$.

The connection between $ \cM_{N}   $ and the 
space of gauge-invariant poynomial functions  of $ Z , Z^{\dagger} $
is a more intricate version of  the connection between $\mR^N/S_N$ 
and symmetric polynomials. Likewise the connection between 
Brauer algebras and gauge-invariant differential operators 
on $ \cM_N$ is a generalization of the connection between 
symmetric groups  and  differential operators 
on  $\mR^N/S_N$.

\vskip.2cm

We present some avenues for future research :  

\begin{enumerate}

\item We wonder if some of these 
{ \bf free particle structures}  
can be obtained from the  { \bf dual supergravity}  side
of AdS/CFT in the sector which is $ SO(4) \times SO(4) $ invariant. 
This would be a non-supersymmetric 
generalization of the LLM \cite{Lin:2004nb} discovery 
of supergravity geometries corresponding to the
free fermions of the holomorphic sector
of the complex matrix model 
\cite{Corley:2001zk,Berenstein:2004kk}.

\item
We have given explicit expressions for 
the free fermion momenta for the $k=0$ sector of 
the $N=2$ matrix theories in terms of the original 
matrix variables. It is an  open problem to find explicit 
expressions for the coordinates of the fermions,
and the wavefunctions as  Slater determinants. 
It is also interesting to explore whether 
this would be useful for the computation of correlators.

\item We have found a {\bf non-holomorphic}  
generalization of the {\bf Calogero-Sutherland} 
Hamiltonian at a fixed coupling.
What is the physics of these non-holomorphic models? 
Can we observe Brauer Algebra wavefunctions in the 
laboratory?

\item  We have presented results on finite $N$ counting 
of complex  matrix model states in terms of Brauer algebras 
at $N=2$. These are related to  reduction 
multiplicities for $ B_{N=2} ( m ,n )$ 
irreps into $ S_m \times S_n$ irreps.  What are these  
{\bf finite $N$ reduction multiplicities} for general 
$m,n,N$, in particular for  $N< m + n$ ?

\item Our analysis has developed
{\bf integrable quantum mechanics models 
for  the space $ \cM_N$} and exploited (Brauer)   
algebras to identify and organise interesting 
spaces of  functions and differential 
operators  on these spaces. $ \cM_N$ is a fibration over 
$ \mR^N/S_N $ which arises in hermitian matrix models
and, like $ \mR^N/S_N $,  has different 
strata where the orbits qualitatively change their 
structure.  While symmetric groups $S_n$ or their inductive 
limit $ S_{ \infty } $ organise functions and differential 
operators on the symmetric product, the Brauer algebras 
$B_N ( m , n ) $ or similarly their 
inductive limit $ B_{ N } ( \infty , \infty ) $ 
organise $ \cM_N$. Results in matrix models, especially multimatrix models
\cite{Donos:2005vm,Kimura:2007wy,Brown:2007xh,BCD,Brown:2008ij,Kimura:2008ac,Bhattacharyya:2008xy,
Kimura:2009jf} can give analogous results for other 
stratified spaces which arise as the space of inequivalent configurations. 
Is it possible to understand the role of algebras,
integrable structures and hidden free particle systems
intrinsically from the
stratified geometries? What is the intrinsic characterization of 
stratified geometries which allow such structures?  
Studies of Hilbert space structures which mirror 
the strata in certain stratified spaces have been done 
\cite{Huebschmann:2007fz}.  
Studying $\cM_N$ from a similar point of view 
 and finding its relations to the Brauer algebra description of 
functions and differential operators would be very interesting.

\item There is a substantial literature discussing 
{\it consistent truncations} of the  Maldacena duality. 
For example, it is known that the $SU(2)$ sector 
defines a consistent truncation to all orders in perturbation 
theory \cite{minahan}. Sectors such as the $ Z , Z^{\dagger} $ sector
are well-defined truncations at zero coupling.
Assuming the strong finite $N$ form of the 
Maladacena conjecture, and making the plausible 
assumption that consistent quantum truncations  of 
a quantum field theory with a string dual have a string dual, 
we are led to ask:  What is the {\bf gauge-string theory dual of 
one free complex matrix} in four dimensions? Similarly
what is the dual of the quantum mechanics from reduction of the 
$ Z , Z^{\dagger}$ sector on $\mR \times S^3$?   
For the large $N$ Gaussian 
Hermitian matrix model (without the quadratic potential) 
there is the non-critical string considered in the old matrix model literature 
\cite{Ginsparg:1993is}. For double scaling limits of the 
complex matrix models, we have the Type-0 string backgrounds 
\cite{Klebanov:2003wg}. For the large $N$ hermitian matrix oscillator
quantum mechanics, which is also a consistent truncation of 
the s-wave sector of  $\cN=4$ SYM in radial quantization,  
there is the proposal \cite{Itzhaki:2004te}. 
A well-known example of a duality
between matrix quantum mechanics and M-theory is given by
\cite{Banks:1996vh}.

\end{enumerate}

We do not have a clear answer to the last question, but the following
remarks are  suggested by the investigations in this paper. 
We conjecture that there exists a string dual of the matrix harmonic oscillator
quantum mechanics discussed in Section \ref{sec:SHO} which has a $2+1$
dimensional space-time and whose physics involves interacting
 strings and branes. 
The $ z_i $  coordinates are positions of $N$ branes 
in $ 2$ space dimensions. By analogy to the treatment in \cite{wittenbound} 
we expect the variables $t_{ij}$ of the Schur decomposition to
describe strings connecting brane $i$ to $j$; here the 
triangular constraint ($t_{ij} = 0$ for $i>j$) will make the dual 
qualitatively different from the standard system of strings and branes 
at weak coupling. This ought to be explained
by an explicit construction of the string theory. 
The Hamiltonian $H$ contains terms $ t { \d \over \d t } $ 
along with $  z_i { \d \over \d z_i} $. Excitations involving polynomials 
in $z_i$ have energies comparable to excitations involving 
$t$. This means that strings and branes have comparable masses.  
Usually string states have masses of order $1$ (with $l_s=1$) 
whereas branes have masses of order $1/g_s$. In this sense, 
the model at hand appears to have $ g_s \sim 1 $. 
An interesting problem is to construct this 
strings and branes model in detail and to provide 
a physical interpretation for the labels of the Brauer algebra 
basis, in particular $k$,  and their constraints at finite $N$.

\vskip1in  

{ \Large 
{ \centerline { \bf Acknowledgements }  } } 

\vskip.2in 
We thank Tom Brown, Robert de Mello Koch,  Diego Rodriguez-Gomez 
for useful discussions and we thank the anonymous referee
for comments which led to an improved exposition. 
SR is supported by an STFC grant ST/G000565/1. 
YK was supported by STFC grant PP/D507323/1.
DT is supported by an STFC studentship.      
YK would like to thank Okayama Institute for Quantum Physics and 
Theoretical Physics Laboratory in RIKEN for hospitality.

\newpage

\begin{appendix}

\section{The Brauer algebra basis}  \label{app:Brauer}

In this appendix we briefly introduce the Brauer algebra basis for gauge invariant polynomials in $Z,\Zd$. 
The Brauer algebra $ B_N ( m , n )$ 
is used to construct a basis of these polynomials of degree
$m$ in $Z$ and degree $n$ in $ Z^{\dagger}$. 
We will not need a precise definition of these algebras here, 
rather we will recall below how their representation theoretic 
data are used to label a useful basis.  
 The definition of these Brauer algebras and their use in constructing 
a basis of gauge invariant operators is found 
 in the original paper \cite{Kimura:2007wy}. A more detailed review of the 
 of the construction  may be found in Section 2 
of \cite{Kimura:2009jf}. 

A Brauer basis operator is a linear combination of multi-trace operators built from $m$ $Z$'s and $n$ $\Zd$'s and is written as
\be
\cO^{\g}_{\a,\b;i,j} (Z,\Zd).
\ee
These operators are constructed by viewing $ Z^{\otimes m } \otimes 
( Z^* )^{\otimes n } $ as operators on $ \vmn$, composing 
them with elements in the Brauer algebra $ Q^{\g }_{ \a , \b ; i , j }  $
and taking a trace \cite{Kimura:2007wy}:
\bea
\cO^{\g}_{\a,\b;i,j} (Z,\Zd) &=& \tr_{m,n} \left( Q^{\gamma}_{\a,\b;i,j} (\mathbf{Z} \otimes \mathbf{Z}^{\ast})
\right)
\eea
The same construction can be done 
with the creation operators of the matrix quantum 
mechanics by replacing $ Z $ with $ A^{\dagger } $ and $ Z^* $ with 
$( B^{\dagger} ) ^T  $ where $T$ denotes matrix transpose. 
These operators diagonalize the two-point function for $ Z , Z^{\dagger} $ 
at zero Yang-Mills coupling or the Fock space inner product for the states created by the $ A^{\dagger} , B^{\dagger} $ of Section \ref{sec:SHO}.

The labels on the operator are as follows:
\begin{enumerate}
\item  $\a , \b$ are Young diagrams with $m$ and $n$ boxes respectively, with $c_1(\a) \le N$ and $c_1(\b) \le N$. They label representations of $U(N)$ as well as $S_m$ and $S_n$ respectively.
\item  $\g = (k, \g_+, \g_-)$ where 
    \begin{enumerate}
 	\item $k$ is an integer in the range $0 \le k \le \min(m,n)$
 	\item $\g_+,\g_-$ are Young diagrams with $m-k$ and $n-k$ boxes respectively, with $c_1(\g_+) + c_1(\g_-)  \le N$.
	\end{enumerate}
	$\g$ labels a representation of the (walled) Brauer algebra $B_N(m,n)$.
\item $i, j$ are indices which run from 1 to the multiplicity $M^{\g}_{\a \b}$ of the representation $(\a,\b)$ of $\mC[S_m \times S_n]$ in the representation $\g$ of the Brauer algebra.
\end{enumerate}

The Brauer representation labelled by $\g = (k, \g_+, \g_-)$ has an associated $U(N)$ composite representation labelled by $\g_c$ which is defined as follows. Using the usual notation in which a Young diagram with row lengths $r_i$ is written $[r_1,r_2,\ldots ,r_N]$, let 
\be
\g_+ = [r_1,r_2,\ldots ,r_p], \qquad \g_- = [s_1,s_2,\ldots ,s_q]
\ee
then providing $p+q \le N$, $\g_c$ is given by
\be \label{eq:gammac}
\g_c =  [r_1,r_2,\ldots ,r_p, 0,0, \ldots, 0, -s_q, -s_{q-1}, \ldots ,-s_1]
\ee
where there are $N-(p+q)$ zeroes inserted. In the language of the mathematics literature $\g_c$ is an $N$-staircase with positive part $\g_+$ and negative part $\g_-$ \cite{Stembridge:1987,Benkart:1992}. 

When we discuss Casimir operators we use the shorthand $C_2(\g)$ for the $U(N)$ quadratic Casimir of the representation $\g_c$, and similarly $r_p(\g)$ or $r_p^{\g}$ for the $p$-th row of $\g_c$. \newline

When $k=0$ the $i,j$ labels are trivial and we have $\alpha = \gamma_+ ,~ 
\beta = \gamma_-$.  Thus $\gamma$ is given by $( k=0 , \a, \b )$ and so the $k=0$ 
operators are thus determined by two Young diagrams $\a$ and $\b$.
To connect with the notation of `coupled representations' in the 
two-dimensional Yang-Mills literature (see e.g. \cite{Gross:1993yt}), we
rename $ \gamma^+ = R , \gamma_- = S $.
If we substitute a unitary matrix in place of $Z$, the $k=0$ polynomials 
coincide with the `coupled characters' $ \chi_{R \bar S } ( U ) $
studied in the context of the string theory 
two-dimensional Yang-Mills \cite{Gross:1993yt}.

At $(m,n) = (1,1)$, suppressing non-essential labels, the Brauer basis is
\bea
\cO^{k=0}_{[1],[\bar{1}]} (Z,\Zd) &=& \tr Z \tr \Zd - \frac{1}{N} \tr Z \Zd \\
\cO^{k=1}_{[1],[\bar{1}]} (Z,\Zd) &=& \frac{1}{N} \tr Z \Zd 
\eea
Here we have suppressed $\g_+$ and $\g_-$ since for a $k=0$ operator it is always the case that $\a = \g_+$ and $\b = \g_-$, and since for the above $k=1$ operator, $\g_+$ and $\g_-$ are both the empty diagram. The multiplicity indices $i,j$ are not relevant for this example.  For further examples of Brauer basis operators see Appendix A.4 of \cite{Kimura:2007wy}.

\section{List of $\g_+$ and $\g_-$ at $N=2$ for given $(m,n)$}  \label{app:gammalist}

Given $(m,n)$ the possible $\g_+$ and $\g_-$ are listed below, along with $\g_c$ as defined in equation (\ref{eq:gammac}). Note that $r_1^{\g}$ (and hence $p_1^{\g}$) distinguishes operators, as does $r_2^{\g}$. We use the shorthand $C_2(\g)$ for the $U(N)$ quadratic Casimir of the representation labelled by $\g_c$. \newline

\textbf{List of $\g_+$ and $\g_-$ when $m \ge n$ using $d = m-n$}
\renewcommand\arraystretch{1.7}  
$$
\begin{array}{cc|c|c|c}
\g_+ & \g_-  &   \g_c   &   k & C_2(\g)  \cr  \hline
[m]  &  [n]  & [m,-n]  &  0  &   m(m+1) + n(n+1)  \cr
[m-1]  &  [n-1]  & [m-1,-(n-1)]  &  1  &   (m-1)(m) + (n-1)(n)  \cr
\vdots  &  \vdots  & \vdots  &  \vdots  &   \vdots  \cr
[d+1]  &  [1]  & [d+1,-1]  &  n-1  &   (d+1)(d+2) + 2  \cr
[d]  &  \emptyset  & [d,0]  &  n  &   d(d+1) \cr
[d-1,1]  &  \emptyset  & [d-1,1]  &  n  &   (d-1)(d)   \cr
\vdots  &  \vdots  & \vdots  &  \vdots  &   \vdots  \cr
\Big[ \,  \ceil{ \frac{d}{2} } , \floor{ \frac{d}{2} } \, \Big]  &  \emptyset  & \Big[ \,  \ceil{ \frac{d}{2} } , \floor{ \frac{d}{2} } \, \Big]  &  n  &  \ceil{ \frac{d}{2} } \Big(  \ceil{ \frac{d}{2} } + 1  \Big) +  \floor{ \frac{d}{2} } \Big( \floor{ \frac{d}{2} } - 1 \Big)   \cr
\end{array}
$$ \newline

\textbf{List of $\g_+$ and $\g_-$ when $m < n$ using $\tilde{d} = n-m$}
$$
\begin{array}{cc|c|c|c}
\g_+ & \g_-  &   \g_c   &   k & C_2(\g)  \cr  \hline
[m]  &  [n]  & [m,-n]  &  0  &   m(m+1) + n(n+1)  \cr
[m-1]  &  [n-1]  & [m-1,-(n-1)]  &  1  &   (m-1)(m) + (n-1)(n)  \cr
\vdots  &  \vdots  & \vdots  &  \vdots  &   \vdots  \cr
[1]  &  [\tilde{d}+1]  & [1,-(\tilde{d}+1)]  &  m-1  &   (\tilde{d}+1)(\tilde{d}+2) + 2  \cr
\emptyset  & [\tilde{d}] & [0,-\tilde{d}]  &  m  &   \tilde{d}(\tilde{d}+1) \cr
\emptyset  & [\tilde{d}-1,1]  & [-1,-(\tilde{d}-1)]  &  m  &   (\tilde{d}-1)(\tilde{d})   \cr
\vdots  &  \vdots  & \vdots  &  \vdots  &   \vdots  \cr
\emptyset & \Big[ \,  \ceil{ \frac{\tilde{d}}{2} } , \floor{ \frac{\tilde{d}}{2} } \, \Big]    & \Big[ \, - \floor{ \frac{\tilde{d}}{2} } , - \ceil{ \frac{\tilde{d}}{2} } \, \Big]  &  m  &  \ceil{ \frac{\tilde{d}}{2} } \Big(  \ceil{ \frac{\tilde{d}}{2} } + 1  \Big) +  \floor{ \frac{\tilde{d}}{2} } \Big( \floor{ \frac{\tilde{d}}{2} } - 1 \Big)   \cr
\end{array}
$$
\renewcommand\arraystretch{1.0}


\section{Brauer counting at large $N$ from Clebsch counting} 
\label{brauerbasiscounting}

In this section we show that $N_{sb}(m,n)$ as defined in (\ref{eq:Nsb}),
\begin{eqnarray} \label{eq:Nsb2}
N_{sb}(m,n) &=& \sum_{\gamma,A} \left( M_{A}^{\gamma} \right)^{2},
\end{eqnarray}
agrees with equation (\ref{eq:Nmn-cl}),
\be \label{eq:Nmn-cl2}
Q_{mt}(m,n) = \sum_{ c_l (1): \sum_{ l} c_l (1) l = m } ~  \sum_{ c_l (2): \sum_{ l} c_l (2) l = n }
~\prod_{l}\frac{ (c_{l}(1)+c_{l}(2 )) ! } {c_{l}(1)!c_{l}(2)!}  .
\ee
We first expand
\begin{eqnarray} 
&& 
 \sum_{\gamma } \sum_{ A }  ( M^{ \gamma  }_ { A } )^2  \cr 
&=& \sum_{k=0}^{\min(m,n)} \sum_{ \gamma_+ \vdash (m-k) }
 \sum_{ \gamma_- \vdash (n-k) } \sum_{ \alpha \vdash m } 
\sum_{  \beta \vdash n  }     
\left( \sum_{ \delta \vdash k }
  g ( \delta , \gamma_+ ; \alpha  ) 
g ( \delta , \gamma_- ; \beta )  \right)^2   \cr
&=&
\sum_{k} \sum_{\gamma_{+},\gamma_{-}} 
\sum_{\alpha,\beta}  
\left( \sum_{ \delta \vdash k }
  g ( \delta , \gamma_+ ; \alpha  ) 
g ( \delta , \gamma_- ; \beta )  \right)  
\left( \sum_{ \delta^{\prime} \vdash k }
  g ( \delta ^{\prime}, \gamma_+ ; \alpha  ) 
g ( \delta^{\prime} , \gamma_- ; \beta )  \right). \label{Nsb}
\end{eqnarray}
Here $g$ is the Littlewood-Richardson coefficient which is defined by
\begin{eqnarray}
g ( \delta , \gamma_+ ; \alpha  ) 
=\frac{1}{k!}\frac{1}{(m-k)!}
\sum_{\sigma_{1}\in S_{k}}\sum_{\sigma_{2}\in S_{m-k}}
\chi_{\delta}(\sigma_{1})\chi_{\gamma_{+}}(\sigma_{2})
\chi_{\alpha}(\sigma_{1} \circ \sigma_{2}).
\end{eqnarray}
To simplify the expresssion (\ref{Nsb}), 
we use the orthogonality of characters 
\begin{eqnarray}
\sum_{R}\chi_{R}(\sigma)\chi_{R}(\tau)=\delta_{T_{\sigma},T_{\tau}}
\frac{n!}{|T_{\sigma}|}
=\delta_{T_{\sigma},T_{\tau}}Sym(T_{\sigma})
\label{orthogonalitycharacter}
\end{eqnarray}
where $T_{\sigma}$ is the size of the conjugacy class 
which contains $\sigma$, and 
$Sym(T_{\sigma})$ represents the number of elements which commute 
with $\sigma$:
\begin{eqnarray}
Sym(T_{\sigma})
&=&c_{1}(\sigma)!1^{c_{1}(\sigma)}
c_{2}(\sigma)!2^{c_{2}(\sigma)} \cdots 
c_{n}(\sigma)!n^{c_{n}(\sigma)} \cr
&=&\prod_{i} c_{i}(\sigma)!i^{c_{i}(\sigma)}
\end{eqnarray}
where $c_{i}(\sigma)$ represents the number of an $i$-cycle in $\sigma$. 
Since $\sigma$ is an element of $S_{n}$, we have 
$\sum_{i=1}^{n}ic_{i}(\sigma)=n$. 

Using (\ref{orthogonalitycharacter}), some factors in (\ref{Nsb}) 
can be rearranged as follows:
\begin{eqnarray}
&&
\sum_{\gamma_{+}}\sum_{\alpha}
g ( \delta , \gamma_+ ; \alpha  ) 
g ( \delta^{\prime} , \gamma_+ ; \alpha  ) \cr
&=&
\sum_{\gamma_{+}}\sum_{\alpha}
\left(\frac{1}{k!(m-k)!}\right)^{2}
\sum_{\sigma_{1},\sigma_{2},\tau_{1},\tau_{2}}
\chi_{\delta}(\sigma_{1})\chi_{\gamma_{+}}(\sigma_{2})
\chi_{\alpha}(\sigma_{1}\circ \sigma_{2})
\chi_{\delta^{\prime}}(\tau_{1})\chi_{\gamma_{+}}(\tau_{2})
\chi_{\alpha}(\tau_{1}\circ \tau_{2}) \cr
&=&
\left(\frac{1}{k!(m-k)!}\right)^{2}
\sum_{\sigma_{1},\sigma_{2},\tau_{1},\tau_{2}}
\chi_{\delta}(\sigma_{1})
\chi_{\delta^{\prime}}(\tau_{1})
\delta_{T_{\sigma_{2}},T_{\tau_{2}}}\frac{(m-k)!}{|T_{\sigma_{2}}|}
\delta_{T_{\sigma_{1}\circ\sigma_{2}},T_{\tau_{1}\circ\tau_{2}}}
\frac{m!}{|T_{\sigma_{1} \circ\sigma_{2}}|} \cr
&=&
\frac{m!}{(k!)^{2}(m-k)!}
\sum_{\tau_{1},\tau_{2}}
\chi_{\delta}(\tau_{1})
\chi_{\delta^{\prime}}(\tau_{1})
\frac{1}{|T_{\tau_{2}}|} 
\frac{1}{|T_{\tau_{1} \circ\tau_{2}}|}|T_{\tau_{1}}||T_{\tau_{2}}|
\cr
&=&
\frac{m!}{(k!)^{2}(m-k)!}
\sum_{\tau_{1},\tau_{2}}
\chi_{\delta}(\tau_{1})
\chi_{\delta^{\prime}}(\tau_{1})
\frac{1}{|T_{\tau_{1} \circ\tau_{2}}|}|T_{\tau_{1}}|
\end{eqnarray}
Then we get the following expression for (\ref{Nsb}) 
\begin{eqnarray} 
 N_{ sb }(m,n)
&=&
\sum_{k=0}^{\min(m,n)}
\frac{m!n!}{(k!)^{4}(m-k)!(n-k)!} \cr
&& 
\sum_{\delta,\delta^{\prime}}
\sum_{\tau_{1},\tau_{2}}
\chi_{\delta}(\tau_{1})
\chi_{\delta^{\prime}}(\tau_{1})
\frac{|T_{\tau_{1}}|}{|T_{\tau_{1} \circ\tau_{2}}|}
\sum_{\rho_{1},\rho_{2}}
\chi_{\delta}(\rho_{1})
\chi_{\delta^{\prime}}(\rho_{1})
\frac{|T_{\rho_{1}}|}{|T_{\rho_{1} \circ\rho_{2}}|} \cr
&=&
\sum_{k}
\begin{pmatrix}
m \\
k
\end{pmatrix}
\begin{pmatrix}
n \\
k
\end{pmatrix} 
\sum_{\tau_{1},\tau_{2}}\sum_{\rho_{1},\rho_{2}}
\delta_{T_{\tau_{1}},T_{\rho_{1}}}
\delta_{T_{\tau_{1}},T_{\rho_{1}}}\frac{1}{|T_{\tau_{1}}|^{2}}  
\frac{|T_{\tau_{1}}|}{|T_{\tau_{1} \circ\tau_{2}}|} 
\frac{|T_{\rho_{1}}|}{|T_{\rho_{1} \circ\rho_{2}}|} \cr
&=&
\sum_{k}
\begin{pmatrix}
m \\
k
\end{pmatrix}
\begin{pmatrix}
n \\
k
\end{pmatrix} 
\sum_{\tau_{1}\in S_{k}}\sum_{\tau_{2}\in S_{m-k}}
\sum_{\rho_{2}\in S_{n-k}}
\frac{1}{|T_{\tau_{1} \circ\tau_{2}}|}
\frac{|T_{\tau_{1}}|}{|T_{\tau_{1} \circ\rho_{2}}|} \cr
&=&
\sum_{k}
\begin{pmatrix}
m \\
k
\end{pmatrix}
\begin{pmatrix}
n \\
k
\end{pmatrix}
\sum_{T_{\tau_{1}}\in S_{k}}\sum_{T_{\tau_{2}}\in S_{m-k}}
\sum_{T_{\rho_{2}}\in S_{n-k}}
|T_{\tau_{1}}|^{2}|T_{\tau_{2}}||T_{\rho_{2}}|
\frac{1}{|T_{\tau_{1} \circ\tau_{2}}|}
\frac{1}{|T_{\tau_{1} \circ\rho_{2}}|} \cr
&=&
\sum_{k}
\sum_{T_{\tau_{1}}\in S_{k}}\sum_{T_{\tau_{2}}\in S_{m-k}}
\sum_{T_{\rho_{2}}\in S_{n-k}}
\frac{
Sym(T_{\tau_{1} \circ\tau_{2}}) } 
{Sym(T_{\tau_{1}}) Sym(T_{\tau_{2}}) } 
\frac{ 
Sym(T_{\tau_{1} \circ\rho_{2}})
}{
Sym(T_{\tau_{1}})Sym(T_{\rho_{2}})
}
\cr
&=&
\sum_{k}
\sum_{T_{\tau_{1}}\in S_{k}}\sum_{T_{\tau_{2}}\in S_{m-k}}
\sum_{T_{\rho_{2}}\in S_{n-k}}
\prod_{l}
\begin{pmatrix}
c_{l}(\tau_{1})+c_{l}(\tau_{2}) \\
c_{l}(\tau_{1})
\end{pmatrix}
\begin{pmatrix}
c_{l}(\tau_{1})+c_{l}(\rho_{2}) \\
c_{l}(\tau_{1})
\end{pmatrix} . \nn
\end{eqnarray}
The above can be rewritten as 
\begin{eqnarray}\label{countkr}   
\sum_{ c_l (1) : \sum{ l c_l (1) = m }}  \sum_{ c_l (2) : \sum { l c_l (2) = n } }
\prod_l         \sum_{k}   \sum_{ c_l(3) :    \sum l c_l(3) = k  }^{ min ( c_l(1) , c_l (2) ) }
{ c_l(1)  !  \over  ( c_l(1) - c_l(3) )!  c_l(3)! }  { c_l(2)  !  \over  ( c_l(2) - c_l(3) )!  c_l(3)! }. \cr
\end{eqnarray} 

We now compare to (\ref{eq:Nmn-cl}), which is the expression
\be \label{eq:app-Nmn-cl2}
Q_{mt}(m,n) = \sum_{ c_l (1): \sum_{ l} c_l (1) l = m } ~  \sum_{ c_l (2): \sum_{ l} c_l (2) l = n }
~\prod_{l}\frac{ (c_{l}(1)+c_{l}(2 )) ! } {c_{l}(1)!c_{l}(2)!}  .
\ee
For any fixed cycle length in $ S_m \times S_n $ consider 
the conjugacy class with $ c_l (1)  $ cycles in $S_m$ and $ c_l(2)$
cycles in $S_n $. 
The factor  $ {  ( c_1 (1) + c_l (2) )!  \over c_l (1) ! c_l (2)! } $ is the number 
of ways of arrangements of   $ ( c_1 (1) + c_l (2) ) $ objects with $c_l(1)$ of one kind (say red) 
and $c_l(2)$ of another kind (say blue).  Suppose we lay our the objects in a line. We can take the first arrangemnet to 
be the one with $c_l(1) $ reds  on the left and $c_l(2)$ blues on the right. Then we permute to generate the rest. 
A general arrangement will have $ c_l(3) $  blues  on the left  among  $c_l(1)-c_l(3) $  red objects  
and $c_l(3) $  reds among $ c_l (2) - c_l(3)   $ blues  on the right.  Of these we have 
 $ {  c_l(1) ! \over k!  ( c_l(1)- c_l(3)  )! } \times {  c_l(2) ! \over k!  ( c_l(2)-c_l(3)  )! }$ arrangements. 
Hence we get 
\begin{eqnarray} 
 {  ( c_1 (1) + c_l (2) )!  \over c_l (1) ! c_l (2)! } = 
\sum_{ c_l(3) =0}^{ min ( c_l(1) , c_l(2) )  }  {  c_l(1) ! \over c_l(3) !  ( c_l(1)- c_l(3)  )! } \times {  c_l(2) ! \over c_l(3) ! 
 ( c_l(2)- c_l(3)  )! }.
\end{eqnarray} 
This proves the desired equality between (\ref{countkr}) and (\ref{eq:app-Nmn-cl2}), and so we conclude that the two countings (\ref{eq:Nmn-cl}) and (\ref{eq:Nsb}) agree:
\bea
N_{sb}(m,n) &=& Q_{mt}(m,n).
\eea


\section{Brauer counting from $GL(N) \times GL(N) \rightarrow  GL(N) $ reduction } 
\label{brauerbasiscounting2}  

We now show that the Brauer basis correctly counts invariants under the adjoint $U(N)$ action
\bea\label{Uact-app}  
Z \rightarrow U Z U^{\dagger} .
\eea  
We consider invariants under (\ref{Uact-app}) constructed 
from objects of the form: 
\begin{eqnarray} 
\T{Z}{i_1}{j_1} \cdots \T{Z}{i_m}{j_m} \T{\Zd}{k_1}{l_1} \cdots  \T{\Zd}{k_n}{l_n}.
\end{eqnarray} 
As far as counting invariants under $U(N)$ action is concerned, the 
problem is equivalent to counting invariants under $GL(N)$. 

The Lie algebra of $GL(N)$ is just the 
full Matrix algebra $M(N, \mathbb{C } ) $ and the symmetric algebra 
over $ S ( M ( N , \mathbb{C } ) ) $ is decomposed into the direct sums 
\cite{Koike:1989}
\begin{eqnarray} 
S  ( M ( N , \mathbb{C } ) )  = \sum_{ \lambda  } 
V_{\lambda , N  } \otimes  ( V_{\lambda , N })^*
\end{eqnarray} 
as $ GL(N ) \otimes GL(N) $ modules. 
The sum is over partitions with length at most $N$, 
i.e. Young diagrams with first column no longer than $N$. 
Restricting to $S_m  ( M ( N , \mathbb{C } ) )$ leads to the 
restriction $| \lambda | =m $, i.e we are looking at the case 
of Young diagrams with $m$ boxes. 

Decomposing the $GL(N) \times GL(N)$ into $GL(N)$ 
we have 
\begin{eqnarray} 
S_m  ( M ( N , \mathbb{C } ) ) = \sum_{\tau ,\eta , \nu , \lambda }
 g ( \tau , \eta   ; \lambda )  g ( \tau  , \nu ; \lambda ) 
                                              V_{\eta, \nu } .
\end{eqnarray}                                         
Here $V_{\eta , \mu } $ is a composite representation
of $GL(N)$.
Therefore the set of invariants 
in $S_m   ( M ( N , \mathbb{C } ) ) \otimes S_n ( M ( N , \mathbb{C } ) )$, 
is
\begin{eqnarray} 
Inv  \left\{   \sum_{\tau ,\eta, \nu ,  \lambda }
 \sum_{\lambda^{\prime} , \tau^{\prime}, \eta^{\prime} , \nu^{\prime }  } 
 g ( \tau , \eta  ; \lambda ) g ( \tau  , \nu ; \lambda )  
 g ( \tau^{\prime}  , \eta^{\prime}    ; \lambda^{\prime} ) 
 g ( \tau^{\prime}  , \nu^{\prime}  ; \lambda^{\prime} )
  V_{\eta, \nu }  \otimes  V_{\eta^{\prime} , \nu^{\prime} }  \right\}
\end{eqnarray} 
which is nonempty only if $ \eta = \nu^{\prime}  $ ,  $ \nu = \eta^{\prime} $. 
Hence the number of invariants is
\begin{eqnarray}\label{symdecmp} 
\sum_{\tau ,\eta, \nu ,\lambda,\tau',\lambda'}  g ( \tau , \eta , \lambda ) g ( \tau , \nu , \lambda ) 
          g ( \tau^{\prime} , \nu , \lambda^{\prime}  )  g ( \tau^{\prime} , \eta , \lambda^{\prime}  ) .
\end{eqnarray} 
We relabel 
\begin{eqnarray} 
&&
\lambda \rightarrow \alpha   \hskip.2in  
\tau \rightarrow \gamma_+  \hskip.2in 
\eta \rightarrow \delta \cr 
&&
\lambda^{\prime} \rightarrow \beta  \hskip.2in  
\tau^{\prime} \rightarrow \gamma_{-} \hskip.2in 
\nu \rightarrow \delta^{\prime} 
\end{eqnarray} 
to get 
\begin{eqnarray} 
&& \sum_{\a,\b,\g_+,\g_-,\delta,\delta'}  g ( \gamma_+ , \delta  ;  \alpha ) g ( \gamma_+  , \delta^{\prime}  , \alpha ) 
          g ( \gamma_- , \delta^{\prime}  , \beta  )  g ( \gamma_-  , \delta  , \beta   )  \cr
&=&  \sum_{\a,\b,\g_+,\g_-} \left( \sum_{ \delta}  g (  \gamma_+ ,  \delta ; \alpha  ) g ( \gamma_-  , \delta  , \beta   ) \right)
\left( \sum_{ \delta'}  g ( \gamma_+  , \delta^{\prime}  , \alpha ) g ( \gamma_- , \delta^{\prime}  , \beta  )  \right)
\end{eqnarray} 
and using the definition of $M^{\gamma}_{\alpha \beta }$ (\ref{eq:multgg}) this becomes
\begin{eqnarray} 
\sum_{\gamma ,  \alpha , \beta } ( M^{\gamma}_{\alpha \beta })^2
\end{eqnarray} 
which is $N_{sb}(m,n)$ from (\ref{eq:Nsb}).

\section{Proofs for $m = n = k$ projectors} \label{app:k=mproofs}
In this appendix,  
we shall show the operator 
(\ref{k=mprojectorinmm}) 
satisfies the following properties:
\begin{eqnarray} \label{projectork=m}
(P^{k=m}_{\alpha})^{2}=P^{k=m}_{\alpha}
\end{eqnarray}
and 
\begin{eqnarray}
\tr_{k,k}(P^{k=m}_{\alpha})=(d_{\alpha})^{2}.       
\label{tracek=mprojector}
\end{eqnarray}
The second equation follows from the Schur-Weyl duality; 
\begin{eqnarray}
V^{\otimes k}\otimes \bar{V}^{\otimes k}
&=&\bigoplus_{\gamma}
V_{\gamma}^{U(N)}\otimes V_{\gamma}^{B_{N}(k,k)}
\cr
&=&
\bigoplus_{\gamma,A}
V_{\gamma}^{U(N)}\otimes
V_{A}^{\mathbb C(S_{k}\times S_{k})}
\otimes V_{\gamma \to A}^{B_{N}(k,k)
\rightarrow \mathbb C(S_{k}\times S_{k})}.
\end{eqnarray}
In the second line, 
we have 
decomposed each irreducible representation $\gamma$ of the Brauer algebra 
into irreducible representations $A$ of the group algebra of 
$S_m \times S_n$. 
Acting with the projector $P^{k=m}_{\alpha}$ 
on this equation and taking a trace 
in $V^{\otimes k}\otimes \bar{V}^{\otimes k}$, we get 
\begin{eqnarray}
\tr_{k,k}(P^{k=m}_{\alpha})=
d_{(\alpha,\alpha)}
=(d_{\alpha})^{2}
\end{eqnarray}
where we have used $Dim\gamma=1$ and $M_{A}^{\gamma}=1$
for $\gamma=(\emptyset,\emptyset,k=m)$. 

The $k$-contraction operator $C_{(k)}$ can be written in many ways, for example
\begin{eqnarray}
C_{(k)}
&=&\sum_{\sigma \in S_k}
C_{\sigma(1)\bar{1}}\cdots 
C_{\sigma(k)\bar{k}}
\cr
&=&
\sum_{\sigma \in S_k}
\sigma
C_{1\bar{1}}\cdots 
C_{k\bar{k}}
\sigma^{-1}
\cr
&=&
\sum_{\bar{\sigma} \in \bar{S}_k}
\bar{\sigma}
C_{1\bar{1}}\cdots 
C_{k\bar{k}}
\bar{\sigma}^{-1}
\end{eqnarray}
The second equality follows from 
\begin{eqnarray}
\sigma C_{i\bar{j}}
=C_{\sigma(i)\bar{j}}\sigma
\end{eqnarray}

In order to show (\ref{projectork=m}), we first calculate 
$(C_{(k)})^{2}$:
\begin{eqnarray}
(C_{(k)})^{2}
&=& \sum_{\rho,\sigma\in S_{k}}
\rho 
C_{1\bar{1}}\cdots C_{k\bar{k}}
\rho^{-1}\sigma
C_{1\bar{1}}\cdots C_{k\bar{k}}
 \sigma^{-1}
 \cr
&=& \sum_{\rho,\sigma\in S_{k}}
\tr_{k}(\rho^{-1}\sigma)
\rho 
C_{1\bar{1}}\cdots C_{k\bar{k}}
 \sigma^{-1}
 \cr
 &=& \sum_{\rho,\sigma\in S_{k}}
N^{C_{\rho^{-1}\sigma}}
\rho 
C_{1\bar{1}}\cdots C_{k\bar{k}}
 \sigma^{-1}
 \cr
  &=& \sum_{\tau,\sigma\in S_{k}}
N^{C_{\tau}}\tau
\sigma
C_{1\bar{1}}\cdots C_{k\bar{k}}
 \sigma^{-1}
 \cr
 &=&
N^{k}\Omega_{k}C_{(k)}
\label{C_k^2}
\end{eqnarray}
where $\Omega_{k}$ is the Omega factor defined by
\begin{eqnarray}
\Omega_{k}=\sum_{\sigma\in S_{k}}N^{C_{\sigma}-k}\sigma
\end{eqnarray}
where $C_{\sigma}$ is the number of cycles in $\sigma$. 
Using the equation (\ref{C_k^2}), we can easily show that 
the projector (\ref{k=mprojectorinmm}) satisfies (\ref{projectork=m}).

We also have another interesting equation for $C_{(k)}$:
\begin{eqnarray}
C_{(k)}p_{\alpha}=C_{(k)}\bar{p}_{\alpha}, 
\end{eqnarray}
which is a consequence of 
\begin{eqnarray}
&&
C_{1\bar{1}}\cdots C_{k\bar{k}}\sigma
=C_{1\bar{1}}\cdots C_{k\bar{k}}\bar{\sigma}^{-1}.
\end{eqnarray}

We finally prove 
(\ref{tracek=mprojector}):
\begin{eqnarray}
\tr_{k,k}(P^{k=m}_{\alpha})
&=&
\frac{d_{\alpha}}{k!Dim\alpha}\tr_{k,k}(C_{(k)}p_{\alpha})
\cr
&=&
\frac{d_{\alpha}}{k!Dim\alpha}
\sum_{\sigma \in S_{k}}
\tr_{k,k}(
\sigma
C_{1\bar{1}}\cdots 
C_{k\bar{k}}
\sigma^{-1}p_{\alpha})
\cr 
&=&
\frac{d_{\alpha}}{Dim\alpha}
\tr_{k,k}(
C_{1\bar{1}}\cdots 
C_{k\bar{k}}
p_{\alpha})
\cr
&=&
\frac{d_{\alpha}}{Dim\alpha}
\tr_{k}(p_{\alpha})
\cr
&=&
\frac{d_{\alpha}}{Dim\alpha}
d_{\alpha}Dim\alpha 
\cr
&=&
(d_{\alpha})^{2}. 
\end{eqnarray}

\section{Constructing an inner product on polynomials}\label{sec:innprod}

The inner product on gauge invariant polynomials $\cO ( Z , \Zd )$ given in (\ref{eq:ZZd-IP}),
\bea
( \cO_1(Z,\Zd) , \cO_2(Z,\Zd) ) &=& \frac{1}{(2\pi)^{N^2}} \int { [dZ d\Zd] \, 
\overline{ \Big( e^{-\sq} \, \cO_1(Z,\Zd) \Big) } 
\Big( e^{-\sq} \, \cO_2(Z,\Zd) \Big) e^{-\tr Z \Zd} } \cr && \label{eq:ZZd-IPapp}
\eea
was introduced by identifying it with the integral representation of the inner
product on matrix harmonic oscillator states $\ket{\Psi}$.

In this appendix we show that this inner product can be derived by
\begin{itemize}
	\item Starting from the inner product arising from the two-point 
function of the zero-dimensional complex matrix model of Ginibre \cite{Ginibre},
\bea   
 (  \cO_1 ( Z , \Zd )  , \cO_2 ( Z  , \Zd )   )_G  
&=& \frac{1}{(2\pi)^{N^2}} \int [ dZ d\Zd ]  \, \overline{\cO_1 ( Z , \Zd )} \cO_2 ( Z , \Zd ) 
e^{-\tr Z \Zd} \cr && \label{integprod}
\eea 
where the normalisation factor is the value of the integral with no insertions;
	\item Requiring $\cH_A$ to be hermitian.
\end{itemize}
We will see that this leads us to the inner product (\ref{eq:ZZd-IPapp}). \newline

The construction proceeds as follows.
We know that $ H_1 , \bar H_1 , H_2 , \bar H_2 , H_L$ 
have eigenstates given by the Brauer basis polynomials 
$ \cO^{ \gamma}_{\alpha \beta } ( Z  , \bar Z )  $
with real eigenvalues. These eigenstates are a complete set of gauge 
invariant polynomials. So in fact any  
inner product diagonal in these labels $ \gamma , \alpha , \beta $  
\bea 
 ( \cO^{ \gamma_1 }_{ \alpha_1 \beta_1 }  ,  \cO^{ \gamma_2 }_{ \alpha_2 \beta_2 } ) = f^{ \gamma_1}_{ \alpha_1 \beta_1 }  \delta^{\gamma_1 \gamma_2 } \delta_{\alpha_1 \alpha_2 } 
\delta_{\beta_1 \beta_2 }
\eea 
for some real $ f^{ \gamma_1}_{ \alpha_1 \beta_1 }$
will guarantee that $ H_1 , \bar H_1 , H_2 , \bar H_2 , H_L$ are hermitian.  

We now give an explicit construction of such an inner product, 
which is also well-defined for general polynomials 
in $Z , Z^{\dagger}$, not just gauge-invariant ones.

The basic idea is to define our inner product on degree $1$ monomials in 
$Z , Z^{\dagger} $ and then extend to arbitrary monomials by Wick's theorem. 
\bea \label{basicpairing}  
( \T{Z}{i}{j} , \T{Z}{k}{l} ) &=& \delta^{ik} \delta^{jl}  \cr 
(  \T{\Zd}{i}{j} , \T{\Zd}{k}{l} ) &=& \delta^{ik}  \delta_{jl }   \cr 
(  \T{Z}{i}{j} , \T{\Zd}{k}{l} ) &=& 0 
\eea 
We generalize to higher degree monomials  
\bea 
 ( \T{Z}{i_1}{j_1} \cdots \T{Z}{i_m}{j_m}  \T{\Zd}{p_1}{q_1} \cdots \T{\Zd}{p_n}{q_n} ,
  \T{Z}{k_1}{l_1} \cdots  \T{Z}{k_m}{l_m}  \T{\Zd}{r_1}{s_1}  \cdots \T{\Zd}{r_n}{s_n} )
\eea 
by summing over different possible pairings 
of the $Z$ on the left with the $Z$ on the right, 
and the $ Z^{\dagger} $ on the left with the $ Z^{\dagger} $ 
on the right, with each individual pairing being given by 
(\ref{basicpairing}).
 { \it Very importantly } we do not 
include contractions between pairs  $Z$ and $ \bar Z $ 
both on the left or both on the right. 
In the Ginibre inner product (\ref{integprod}), 
we have $ ( 1 , Z \bar Z ) \ne 0 $ which shows the inner product under construction 
is different to (\ref{integprod}).

To prove $G_2$ is hermitian with this inner product we first work with the basic pairing.
\bea \label{basicherm}  
( \T{Z}{i}{j} , \T{(G_2)}{p}{q}  \T{Z}{k}{l} ) &=&  
  \T{\delta}{k}{q} ( \T{Z}{i}{j} ,  \T{Z}{p}{l} ) ~=~ 
  \T{\delta}{k}{q} \delta^{ip} \delta_{jl}  \cr 
( \T{(G_2)}{p}{q}  \T{Z}{i}{j} ,  \T{Z}{k}{l} ) &=&
  \T{\delta}{i}{q}  ( \T{Z}{p}{j} ,  \T{Z}{k}{l} ) ~=~  
  \T{\delta}{i}{q} \delta^{pk} \delta_{jl} 
\eea 
We thus find, on these degree $1$ monomials 
\be
(\T{(G_2)}{p}{q})^{h} = \T{(G_2)}{q}{p},
\ee
where $h$ denotes hermitian conjugate.
When we consider the action of $\T{(G_2)}{p}{q} $ on a general pairing 
\bea 
 ( \T{Z}{i_1}{j_1} \cdots \T{Z}{i_m}{j_m}  \T{\Zd}{p_1}{q_1} \cdots \T{\Zd}{p_n}{q_n} ,
  \T{(G_2)}{p}{q}  \T{Z}{k_1}{l_1} \cdots  \T{Z}{k_m}{l_m}  \T{\Zd}{r_1}{s_1}  \cdots \T{\Zd}{r_n}{s_n} )
\eea 
we can use the fact that $\T{(G_2)}{p}{q}$ acts as a 
derivation, so that the right factor becomes a sum 
of terms with  the $\T{(G_2)}{p}{q}$ acting on each successive 
$ Z $ or $\Zd $. The action on $\Zd$ gives zero. 
For each term in this sum, the inner product is 
a sum over Wick contractions. For each Wick contraction 
of the form 
\be
( Z , G Z ) ( Z , Z ) \cdots (\Zd , \Zd ) \cdots 
\ee
we can move the $\T{(G_2)}{p}{q}$ over to the left to give 
$\T{(G_2)}{q}{p}$ using (\ref{basicherm}). 
We can recollect the sum over Wick contractions to get 
\bea 
( \T{(G_2)}{q}{p} \T{Z}{i_1}{j_1} \cdots \T{Z}{i_m}{j_m}  \T{\Zd}{p_1}{q_1} \cdots \T{\Zd}{p_n}{q_n} ,
  \T{Z}{k_1}{l_1} \cdots  \T{Z}{k_m}{l_m}  \T{\Zd}{r_1}{s_1}  \cdots \T{\Zd}{r_n}{s_n} ).
\eea 
This establishes for any monomial in $ Z ,\Zd$ 
that   $ (\T{(G_2)}{p}{q})^{h} = \T{(G_2)}{q}{p}$, and by 
linearity this extends to any polynomial.  Having established 
\bea \label{eq:G2herm}
(\T{(G_2)}{p}{q})^{h} &=& \T{(G_2)}{q}{p}
\eea 
it easily follows that
\bea 
( \tr G_2 )^{h}  = \tr G_2  \qquad  \mathrm{and} \qquad ( \tr G_2^2 )^{h} = \tr G_2^2  
\eea 
and similarly we find 
\bea 
( \T{(G_3)}{i}{j} )^{h} &=& \T{(G_3)}{j}{i}  \cr
( \tr G_3 )^{h}  &=& \tr G_3 \cr 
(  \tr G_3^2 )^{h} &=& \tr G_3^2 \cr 
( \tr ( G_2 G_3 ))^h   &=& \tr ( G_3 G_2 ) ~=~ \tr ( G_2 G_3 ) 
\eea 
where the last equality follows since the entries of $G_2$ and $G_3$ commute.

We can also derive the above relations by noting that
\bea
(\T{Z}{i}{j})^h &=& \left( \dbyd{Z} \right)^{\!\! j}_i , \cr \qquad (\T{\Zd}{i}{j})^h &=& \left( \dbyd{\Zd} \right)^{\!\! j}_i \cr
\Rightarrow \left( \T{(G_2)}{p}{q} \right)^{h} &=&  \left( \T{Z}{p}{i} \left( \dbyd{Z} \right)^{\!\! i}_q \right)^h 
                                  ~=~  \T{Z}{q}{i} \left( \dbyd{Z} \right)^{\!\! i}_p  ~=~ \T{(G_2)}{q}{p}
\eea
and similarly for $G_3$ etc.

The above proofs work by construction since we have defined our inner product to 
have the same properties as the oscillator inner product 
for $A^{\dagger} B^{\dagger }$ and exploited the similarities
\bea
\T{(G_2)}{i}{j}  \simeq \T{( A^{\dagger } A )}{i}{j},  &&
\T{(G_3)}{i}{j}  \simeq  \T{( B^{\dagger } B )}{i}{j}. 
\eea

We now derive an integral form of this inner product. We begin with
(\ref{integprod}) and normal order by removing all contributions to 
the inner product from self-contractions in the wavefunctions. 
It was observed in (\ref{eq:lapwick}) that the laplacian 
generates Wick contractions so we define (c.f.~\cite{Polchinski:1998ordering})
\bea\label{modinprod}  
: \cO ( Z , \Zd ) : &=& ( 1 - \Box + { \Box^2\over 2 } + \cdots ) \cO ( Z ,\Zd )
~=~ e^{ - \Box }   \cO ( Z , \Zd )
\eea 
and our inner product becomes the following modification of (\ref{integprod}):
\bea
( \cO_1(Z,\Zd) , \cO_2(Z,\Zd) ) &=& \big( : \cO_1(Z,\Zd): \, , \,  : \cO_2(Z,\Zd) : \big)_G   \cr
&=& \frac{1}{(2\pi)^{N^2}} \int { [dZ d\Zd] \, 
\overline{ \Big( e^{-\sq} \, \cO_1(Z,\Zd) \Big) } 
\Big( e^{-\sq} \, \cO_2(Z,\Zd) \Big) e^{-\tr Z \Zd} } \cr && \label{eq:ZZd-IPappend}
\eea
which is (\ref{eq:ZZd-IPapp}) as we set out to show.

\end{appendix} 

\thispagestyle{empty}


\newpage
\bibliographystyle{JHEP}    	
\bibliography{FBCrefs}	    	

\end{document}